\documentclass[aps,pra,onecolumn,showpacs,superscriptaddress]{revtex4}
\usepackage{graphicx}
\usepackage{amsmath}
\usepackage{amssymb}
\usepackage{lscape}

\input{epsf}

\begin{document}

\title{Three $s$-wave interacting fermions 
under anisotropic harmonic confinement: 
Dimensional crossover of energetics and virial coefficients}

\author{Seyed Ebrahim Gharashi}
\affiliation{Department of Physics and Astronomy,
Washington State University,
  Pullman, Washington 99164-2814, USA}
\author{K.~M.~Daily}
\affiliation{Department of Physics and Astronomy,
Washington State University,
  Pullman, Washington 99164-2814, USA}
\author{D. Blume}
\affiliation{Department of Physics and Astronomy,
Washington State University,
  Pullman, Washington 99164-2814, USA}
\affiliation{ITAMP, Harvard-Smithsonian Center for Astrophysics,
60 Garden Street, Cambridge, Massachusetts 02138, USA}

\date{\today}

\begin{abstract}
We present essentially exact solutions of the Schr\"odinger equation 
for three fermions in two different spin states with 
zero-range $s$-wave interactions 
under harmonic confinement. 
Our approach covers
spherically symmetric, 
strictly two-dimensional,
strictly one-dimensional, 
cigar-shaped, and
pancake-shaped traps.
In particular,
we discuss the transition from quasi-one-dimensional 
to strictly one-dimensional and from quasi-two-dimensional 
to strictly two-dimensional geometries.
We determine and interpret
the eigenenergies of the system 
 as a 
function of the trap geometry
and the strength of the zero-range interactions.
The eigenenergies are used to investigate
the dependence of the second- and third-order virial coefficients, 
which play an important role in the virial expansion 
of the thermodynamic potential,
 on the geometry of the trap.
We show that the second- and third-order
virial coefficients for anisotropic confinement geometries
are, for experimentally relevant temperatures, 
very well approximated by those for the spherically symmetric
confinement for all $s$-wave scattering lengths.
\end{abstract}

\maketitle

\section{Introduction}
\label{sec_introduction}

There has been extensive interest in ultracold atom
physics in the last decade~\cite{review_blume,review_giorgini,review_bloch}.
Ultracold atomic bosonic and fermionic gases are realized experimentally 
under varying external confinements.
In these experiments, the number of particles and the scattering length of 
the two-body interactions are tunable~\cite{chin_rmp}.
Although the complete energy spectrum of the many-body system
cannot, in general, be obtained from first principles, the energy 
spectra of selected few-body systems can, in some cases, be 
determined within a microscopic quantum mechanical 
framework~\cite{busch,calarco-r,calarco,kestner,blume-greene_prl,javier-greene_prl}.
In some cases, the properties of the few-body system have then been 
used to predict
the properties of the corresponding many-body 
system~\cite{ho-1,ho-2,rupak,drummond_prl,drummond_pra,drummond_2d,salomon-2010,zwierlein-2012,daily_2012}.

The behavior of atomic and molecular systems depends strongly 
on the dimensionality of the 
system~\cite{olshanii1,olshanii2,petrov,esslinger}.
In three dimensions, e.g., weakly-bound two-body $s$-wave states exist 
when the $s$-wave scattering length is large and positive but not when 
it is negative.
In strictly one- and two-dimensional geometries, in contrast, $s$-wave 
bound states exist for all values of the $s$-wave 
scattering length~\cite{busch}.

In ultracold atomic gases, the de Broglie wavelength of the atoms
is much larger than the van der Waals length that
characterizes the two-body interactions.
This allows one to replace the van der Waals interaction potential
in free-space low-energy scattering calculations
by a zero-range $s$-wave 
pseudopotential~\cite{fermi,huang,huang2}.
If the particles are placed in an external trap, 
the validity of the pseudopotential treatment (at least if
implemented without accounting for the energy-dependence
of the coupling strength) requires 
that the van der Waals length is much smaller than the 
characteristic trap length~\cite{blume_greene_pra,bolda_pra}.
In many cases, the use of pseudopotentials greatly simplifies the 
theoretical treatment. For example, the eigenequation for two 
particles interacting
through 
a $s$-wave pseudopotential
under harmonic confinement has been derived
analytically for spherically symmetric, strictly one-dimensional, 
strictly two-dimensional and anisotropic harmonic 
potentials~\cite{busch,calarco-r,calarco}.

The $s$-wave pseudopotential has also been applied successfully 
to a wide range of three-body problems, either in free space 
or under 
confinement~\cite{nielsen,kartavtsev,rittenhouse-2010,mora,mora-2005,werner,petrov3}. 
The present paper develops an efficient numerical framework for 
treating the three-body system under anisotropic harmonic confinement. 
The developed formalism allows us to study the dependence of the 
three-body properties on the dimensionality of the system. 
We focus on fermionic systems consisting of two identical 
spin-up atoms and one spin-down atom. 
The dimensional crossover of two-component Fermi gases has 
attracted a great deal of interest 
recently~\cite{sommer-2012, kohl-2012, thomas-2012}. 
This paper considers the three-body 
analog within a microscopic quantum mechanical framework.
We note that our framework readily generalizes to bosonic 
three-body systems. The study of the dimensional crossover 
of bosonic systems is interesting as it allows one to study 
how, under experimentally realizable conditions, Efimov 
trimers~\cite{braaten} that are known to exist in three-dimensional 
space disappear as the confinement geometry is tuned to an effectively 
low-dimensional geometry~\cite{nishidatan}.

This paper generalizes the methods developed in Refs.~\cite{mora,kestner}
for three equal-mass fermions in two different pseudospin states
under spherically symmetric harmonic confinement to anisotropic
harmonic confinement.
We develop an efficient and highly accurate algorithm to calculate the
eigenenergies and eigenstates of the system up to relatively
 high energies as functions of the interaction strength and 
aspect ratio of the trap.
Several applications are considered:
{\em{(i)}} The BCS-BEC crossover curve is analyzed throughout
the dimensional crossover.
{\em{(ii)}} For large and small aspect ratios, the energy spectra 
are analyzed in terms of strictly one-dimensional and
strictly two-dimensional effective three-body Hamiltonian.
{\em{(iii)}} The second- and 
third-order virial coefficients are analyzed as functions
of the temperature, aspect ratio and scattering length.
In particular, we show that the high-temperature limit of the
third-order virial coefficient $b_3$ at unitarity is independent of 
the shape of the trap in agreement with expectations 
derived through use of the local density approximation.
For finite scattering lengths, $b_2$ and $b_3$ for anisotropic
harmonic confinement are well approximated by those
for isotropic harmonic confinement.

The remainder of this paper is organized as follows. 
Section~\ref{sec_formalsolution} presents a formal solution to the
problem of three $s$-wave interacting fermions confined in an
axially symmetric harmonic trap.
We also consider the extreme cases of strictly one-dimensional and
strictly two-dimensional confinement.
Sections~\ref{sec_cigar} and~\ref{sec_pancake} apply 
the formal solution to
cigar-shaped and pancake-shaped traps, respectively.
We determine a large portion of the eigenspectrum
as a function of the scattering length and discuss 
the transition to strictly one-dimensional and
strictly two-dimensional geometries. 
Section~\ref{sec_virial_coeff} uses the two- and three-body 
eigenspectra 
to calculate the second- and third-order virial
coefficients as a function of the temperature and 
the geometry of the confinement.
Finally, Sec.~\ref{sec_conclusion} concludes.


\section{Formal solution}
\label{sec_formalsolution}

We consider a two-component Fermi gas consisting of two spin-up
atoms and one spin-down atom with interspecies $s$-wave
 interactions under anisotropic harmonic confinement.
We refer to the two spin-up atoms as particles 1 and 2,
 and to the spin-down atom as particle 3.
We introduce the single-particle Hamiltonian
 $H_0({\bf{r}}_j,\mathcal{M})$ 
for the $j^{th}$ particle with mass $\mathcal{M}$ under harmonic confinement,
\begin{equation}
\label{eq_H0}
H_0({\bf{r}}_j,\mathcal{M})=\frac{-\hbar^2}{2\mathcal{M}}{\boldsymbol{\nabla}}^2_{{\bf{r}}_j}+\frac{1}{2}\mathcal{M} (\omega^2_zz_j^2 + \omega^2_{\rho} \rho_j^2).
\end{equation}
Here, ${\bf{r}}_j$ is measured with respect to the trap center, and 
in cylindrical coordinates we have
 ${\bf{r}}_j=(z_j, \rho_j, \phi_j)$.
In Eq.~(\ref{eq_H0}),
 $\omega_z$ and $\omega_{\rho}$
 are the angular trapping frequencies in the $z$- 
and $\rho$-directions, respectively. 
The aspect ratio $\eta$ of the trap is defined through $\eta = \omega_{\rho} / \omega_z$.
In this paper, we consider cigar-shaped traps with $\eta>1$
as well as pancake-shaped traps with $\eta<1$.
Our three-particle Hamiltonian $H$ then reads
\begin{equation}
H = \sum_{j=1}^3 H_0({\bf{r}}_{j}, \mathcal{M}) +V_{\rm{int}},
\label{eq_Hamiltonian}
\end{equation}
where $V_{\rm{int}}$ accounts for the interspecies $s$-wave two-body interactions,
\begin{equation}
\label{eq_v_int}
V_{\rm{int}} = V_{\rm{ps}}^{\rm{3D}}({{\bf{r}}_{31}}) + V_{\rm{ps}}^{\rm{3D}}({{\bf{r}}_{32}}).
\end{equation}
The regularized pseudopotential $V_{\rm{ps}}^{\rm{3D}}$ is characterized 
by the three-dimensional $s$-wave scattering length $a^{\rm{3D}}$~\cite{fermi, huang, huang2},
\begin{equation}
\label{eq_pseudopotential}
V_{\rm{ps}}^{\rm{3D}}({{\bf{r}}_{jk}}) = \frac{4 \pi \hbar^2 a^{\rm{3D}}}{\mathcal{M}} \delta ({\bf{r}}_{jk})\frac{\partial}{\partial r_{jk}}r_{jk},
\end{equation}
where ${{\bf{r}}_{jk}}={{\bf{r}}_{j}}-{{\bf{r}}_{k}}$ and
$r_{jk}=|{\bf{r}}_{jk}|$.

 Since the trapping potential is quadratic,
the relative and center of mass degrees of freedom 
separate and we rewrite the Hamiltonian $H$ in terms of 
the relative Hamiltonian $H_{\rm{rel}}$ and the center of mass
Hamiltonian $H_{\rm{cm}}$,
 $H = H_{\rm{rel}} + H_{\rm{cm}}$. 
In the following, we obtain solutions
 to the relative three-body Schr\"odinger equation 
$H_{\rm{rel}} \Psi=E_{\rm{3b}} \Psi$,
where
\begin {eqnarray}
\label{eq_rel_Hamiltonian}
{H_{\rm{rel}}}= H_{\rm{rel},0} + V_{\rm{int}}
\end {eqnarray}
with 
\begin {eqnarray}
\label{eq_rel_Hamiltonian_0}
H_{\rm{rel},0}=H_0({\bf{r}},\mu) + H_0({\bf{R}},\mu).
\end {eqnarray}
In Eq.~(\ref{eq_rel_Hamiltonian_0}), $\mu$ is the two-body reduced mass,
 $\mu=\mathcal{M}/2$, 
and the relative Jacobi coordinates ${\bf{r}}$ and ${\bf{R}}$
 are defined through 
${\bf{r}}={\bf{r}}_{31}$
and
${\bf{R}}=\frac{2}{\sqrt{3}} (\frac{{{\bf{r}}_{1}}+
{{\bf{r}}_{3}}}{2}-{\bf{r}}_{2})$.
Depending on the context, we use either  ${\bf{r}}$ and ${\bf{R}}$ or 
${\bf{r}}_{31}$ and ${\bf{r}}_{32}$ to describe the relative 
degrees of freedom of the three-body system.

To determine the relative three-body
wave function $\Psi({\bf{r}},{\bf{R}})$,
we take advantage of the fact that the solutions to the ``unperturbed''
relative Hamiltonian $H_{\rm{rel},0}$ are known and consider the
Lippmann-Schwinger equation (see, e.g., Ref.~\cite{kestner})
\begin{eqnarray}
\label{eq_lippmann-schwinger}
\Psi({\bf{r}},{\bf{R}})= - \int  G(E_{\rm{3b}};{\bf{r}},{\bf{R}};{\bf{r'}},{\bf{R'}})
V_{\rm{int}}({\bf{r'}},{\bf{R'}})
\Psi({\bf{r'}},{\bf{R'}})~d{\bf{r'}}d{\bf{R'}}.
\end{eqnarray}
The Green's function $G$ for the two ``pseudoparticles'' of mass 
$\mu$ associated with the Jacobi vectors ${\bf{r}}$ and ${\bf{R}}$
is defined in terms of the eigenstates  
$\Phi_{\boldsymbol\lambda_1}({\bf{r}}) \Phi_{\boldsymbol\lambda_2}({\bf{R}})$
and the eigenenergies
$E_{\boldsymbol\lambda_1}+E_{\boldsymbol\lambda_2}$ of $H_{\rm{rel},0}$,
\begin {eqnarray}
\label{eq_two_Green}
 G(E_{\rm{3b}};{\bf{r}},{\bf{R}};{\bf{r'}},{\bf{R'}})=
\sum_{{\boldsymbol\lambda_1},{\boldsymbol\lambda_2}}
\frac{
\Phi_{\boldsymbol\lambda_1}^*({\bf{r'}}) \Phi_{\boldsymbol\lambda_2}^*({\bf{R'}})
\Phi_{\boldsymbol\lambda_1}({\bf{r}}) \Phi_{\boldsymbol\lambda_2}({\bf{R}})
}{
(E_{\boldsymbol\lambda_1}+E_{\boldsymbol\lambda_2})-E_{\rm{3b}}
}.
\end {eqnarray}
Here, 
 ${\boldsymbol\lambda}$ 
collectively denotes the quantum numbers needed to 
label the single-particle harmonic osillator states.
In cylindrical coordinates, we have 
${\boldsymbol\lambda}=(n_z,n_{\rho},m)$
with $n_z = 0, 1, 2, \cdots$,
$n_{\rho} = 0, 1, 2, \cdots$, and
$m = 0, \pm 1, \pm2, \cdots$.
The single-particle harmonic oscillator eigenenergies and eigenstates read
\begin{equation}
\label{eq_ond-b_energy}
E_{\boldsymbol \lambda}=
\left(n_z+\frac{1}{2}\right) \hbar \omega_z + 
\left(2n_{\rho}+|m|+1\right)\eta \hbar \omega_z
\end {equation}
and
\begin{equation}
\label{eq_ond-b_wavefunction}
\Phi_{\boldsymbol\lambda}({\bf{r}})= 
\varphi_{n_z}(z)R_{n_{\rho},m}(\rho)\frac{e^{i m \phi}}{\sqrt{2 \pi}},
\end {equation}
where
\begin{equation}
\label{eq_one-d_wavefunction}
\varphi_{n_z}(z)=
\sqrt{\frac{1}{a_z \sqrt{\pi}~2^{n_z} ~n_z!}}
\exp{\left(-\frac{z^2}{2 a_z^2}\right)} H_{n_z}(z/a_z)
\end {equation}
and
\begin{equation}
\label{eq_two-d_wavefunction}
R_{n_{\rho},m}(\rho)=\sqrt{\frac{2 ~ \eta ~{n_{\rho}!}}{a_z^2(n_{\rho}+ |m|)!}} \exp{\left(-\frac{ \eta \rho^2}{2 a_z^2}\right)}\left(\frac{\eta^{1/2} \rho}{a_z}\right)^{|m|} 
L_{n_\rho}^{(|m|)}\left(\eta \rho^2/a_z^2\right).
\end {equation}
In the last two equations, $H_{n_z}(z/a_z)$ and $L_{n_{\rho}}^{(|m|)}\left(\eta \rho^2/a_z^2\right)$ 
denote Hermite and associated Laguerre polynomials, respectively.
Throughout most of Secs.~\ref{sec_formalsolution}-\ref{sec_pancake}, we
use the oscillator energy $E_z$ and oscillator
length $a_z$ [$E_z=\hbar \omega_z$
and $a_z=\sqrt{\hbar/(\mu \omega_z)}$] as our energy and length units.

In Eqs.~(\ref{eq_lippmann-schwinger})-(\ref{eq_two-d_wavefunction}), 
we employ cylindrical coordinates 
since this choice allows us to write the 
Green's function $G$ compactly.
 However, the two-body $s$-wave interaction potential is
most conveniently expressed in 
spherical coordinates [see Eq.~(\ref{eq_pseudopotential})].
Since the pseudopotential $V_{\rm{ps}}^{\rm{3D}}({\bf{r}})$ acts 
only at a single point, namely at $r=0$,
it imposes a boundary condition on the relative three-body wave function
$\Psi({\bf{r}},{\bf{R}})$ (see, e.g., Ref.~\cite{petrov3}),
\begin{equation}
\label{eq_boundary}
\left. \Psi({\bf{r}},{\bf{R}})\right|_{r \rightarrow 0} 
\approx 
\frac{f({\bf{R}})}{4 \pi a_z^{3/2}}\left(\frac{a_z}{r} - \frac{a_z}{a^{\rm{3D}}}\right).
\end {equation}
The unknown function $f({\bf{R}})$ can be interpreted 
as the relative wave function of the center of mass 
of the interacting pair and the third particle.
Similarly, the pseudopotential $V_{\rm{ps}}^{\rm{3D}}({\bf{r}}_{32})$
imposes a boundary condition on
the wave function  $\Psi({\bf{r}},{\bf{R}})$
when $r_{32}\rightarrow 0$.
Since the wave function $\Psi({\bf{r}},{\bf{R}})$
must be anti-symmetric under the exchange 
of the two identical fermions, i.e.,
$P_{12}\Psi({\bf{r}},{\bf{R}}) = -\Psi({\bf{r}},{\bf{R}})$,
where $P_{12}$ exchanges particles 1 and 2,
the properly anti-symmetrized boundary condition 
 corresponding to $V_{\rm{ps}}^{\rm{3D}}({\bf{r}}_{32})$ reads
\begin{equation}
\label{eq_boundary2}
\left. \Psi({\bf{r}}_{32},{\bf{R}}_{32})\right|_{r_{32} \rightarrow 0} 
\approx -\frac{f({\bf{R}}_{32})}{4 \pi a_z^{3/2}}
\left(\frac{a_z}{r_{32}} - \frac{a_z}{a^{\rm{3D}}}\right).
\end{equation}
Here, we defined
${\bf{R}}_{32} =
\frac{2}{\sqrt{3}}\left(\frac{{\bf{r}}_2+{\bf{r}}_3}{2}-{\bf{r}}_{1}\right).$

To simplify the right hand side of Eq.~(\ref{eq_lippmann-schwinger}),
we impose the limiting behaviors of $\Psi({\bf{r'}},{\bf{R'}})$ for 
$r'_{31}\rightarrow 0$ 
and
$r'_{32}\rightarrow 0$,
and expand $f({\bf{R'}})$ in terms of the non-interacting harmonic 
oscillator functions,
$f({\bf{R'}})= \sum_{\boldsymbol\lambda'}f_{\boldsymbol\lambda'} \Phi_{\boldsymbol\lambda'}({\bf{R'}})$.
Using Eq.~(\ref{eq_two_Green}) for $G$
and orthonormality of the single-particle
harmonic oscillator functions, we find
\begin{eqnarray}
\label{eq_wf_expansion}
 \Psi({\bf{r}},{\bf{R}})=\frac{E_z a_z^{3/2}}{2}\sum_{\boldsymbol\lambda}f_{\boldsymbol\lambda}
\left[
\mathcal{G}^{\rm{3D}}\left(E_{\rm{3b}}-E_{\boldsymbol\lambda};{\bf{r}};{\bf{0}}\right) \Phi_{\boldsymbol\lambda}({\bf{R}}) -
\mathcal{G}^{\rm{3D}}\left(E_{\rm{3b}}-E_{\boldsymbol\lambda};\frac{{\bf{r}} + \sqrt{3} {\bf{R}}}{2};{\bf{0}}\right)
 \Phi_{\boldsymbol\lambda}\left(\frac{\sqrt{3}{\bf{r}} - {\bf{R}}}{2}\right)
\right].
\end {eqnarray}
Here, we used that ${\bf{r}}_{32}$ can be written as 
$({\bf{r}} + \sqrt{3}{\bf{R}})/2$
and introduced the one-body Green's function
$\mathcal{G}^{\rm{3D}}\left(E;{\bf{r}};{\bf{r'}}\right)$
for the pseudoparticle of mass $\mu$ that is associated with the relative
distance vector ${\bf{r'}}$,
\begin{eqnarray}
\label{eq_one_green}
\mathcal{G}^{\rm{3D}}\left(E;{\bf{r}};{\bf{r'}}\right) = 
\sum_{\boldsymbol\lambda'}\frac{\Phi_{\boldsymbol\lambda'}^*({\bf{r'}})
\Phi_{\boldsymbol\lambda'}({\bf{r}})
}{E_{\boldsymbol\lambda'}-E}.
\end {eqnarray}
The one-body Green's function
$\mathcal{G}^{\rm{3D}}\left(E_{\rm{2b}};{\bf{r}};{\bf{r'}}\right)$ 
with ${\bf{r'}}={\bf{0}}$
coincides with the solution to
the relative Schr\"odinger equation 
for two particles under harmonic confinement
interacting through the zero-range pseudopotential
$V_{\rm{ps}}^{\rm{3D}}({\bf{r}})$ with $s$-wave 
scattering length $a^{\rm{3D}}$ and relative two-body
energy $E_{\rm{2b}}$.
$\mathcal{G}^{\rm{3D}}\left(E;{\bf{r}};{\bf{0}}\right)$  is known for 
all aspect ratios $\eta$~\cite{calarco-r,calarco}
(see also Secs.~{\ref{sec_cigar}} and~{\ref{sec_pancake}}).

To determine the expansion coefficients $f_{\boldsymbol\lambda}$,
we apply the operation 
$\left. \frac{\partial}{\partial r}(r \cdot )\right|_{r \rightarrow 0}$
to the left hand side and the right hand side of 
Eq.~(\ref{eq_wf_expansion}), i.e., we multiply both sides 
of Eq.~(\ref{eq_wf_expansion}) by $r$, then apply the
derivative operator and lastly take the limit $r \to 0$.
Defining
\begin{align}
\label{eq_f}
\mathcal{F}^{\rm{3D}}(\epsilon_{\boldsymbol{\lambda}},\eta)= 2 \pi E_z a_z^3
\left. \frac{\partial}{\partial r}\left\{r\mathcal{G}^{\rm{3D}}\left([\epsilon_{\boldsymbol{\lambda}}+\eta+1/2]E_z;{\bf{r}};{\bf{0}}\right)\right\}\right|_{r \rightarrow 0}
\end{align}
with
$(\epsilon_{\boldsymbol{\lambda}}+\eta + 1/2)E_z= 
E_{\rm{3b}}-E_{\boldsymbol{\lambda}}$, 
we find
\begin{align}
\label{eq_eigen1}
& -\frac{a_z}{2 \pi a^{\rm{3D}}}\sum_{\boldsymbol{\lambda'}}
f_{\boldsymbol{\lambda'}}\Phi_{\boldsymbol\lambda'}({\bf{R}})=
\nonumber\\
& 
\sum_{\boldsymbol{\lambda'}}f_{\boldsymbol{\lambda'}}
\left\{\frac{1}{2 \pi}\mathcal{F}^{\rm{3D}}(\epsilon_{\boldsymbol{\lambda'}},\eta)
\Phi_{\boldsymbol\lambda'}\left({\bf{R}}\right)-
E_z a_z^3 \mathcal{G}^{\rm{3D}}\left([\epsilon_{\boldsymbol{\lambda'}}+\eta+1/2]E_z;\frac{\sqrt{3}}{2}{\bf{R}};{\bf{0}}\right)
\Phi_{\boldsymbol\lambda'}\left(\frac{-{\bf{R}}}{2}\right)\right\}.
\end{align}
The quantity $\epsilon_{\boldsymbol{\lambda'}}$
can be interpreted as a non-integer quantum number 
associated with the interacting pair.
If we multiply Eq.~(\ref{eq_eigen1}) by $\Phi^*_{\boldsymbol\lambda}({\bf{R}})$
and integrate over ${\bf{R}}$, we find an implicit eigenequation 
for the relative three-body energy $E_{\rm{3b}}$ or equivalently, the non-integer
quantum number $\epsilon_{\boldsymbol{\lambda}}$,
\begin{eqnarray}
\label{eq_eigen2}
\sum_{\boldsymbol{\lambda'}}\left[
I_{{\boldsymbol{\lambda}},\boldsymbol{\lambda'}}^{\rm{3D}}
\left(\epsilon_{\boldsymbol{\lambda'}}\right)-
\mathcal{F}^{\rm{3D}}\left(\epsilon_{\boldsymbol{\lambda}},\eta\right) 
\delta_{{\boldsymbol{\lambda}},\boldsymbol{\lambda'}}\right]
f_{\boldsymbol{\lambda'}}=
\frac{a_z}{a^{\rm{3D}}}f_{\boldsymbol{\lambda}},
\end{eqnarray}
where
\begin{eqnarray}
\label{eq_Ilambda}
I_{{\boldsymbol{\lambda}},\boldsymbol{\lambda'}}^{\rm{3D}}
\left(\epsilon_{\boldsymbol{\lambda'}}\right) 
= 2 \pi E_z a_z^3 
\int
\mathcal{G}^{\rm{3D}}
\left(
[\epsilon_{\boldsymbol{\lambda'}}+\eta+1/2]E_z;
\frac{\sqrt{3}}{2}{\bf{R}};{\bf{0}}\right) 
\Phi_{\boldsymbol\lambda'}\left(\frac{-{\bf{R}}}{2}\right) 
\Phi_{\boldsymbol\lambda}^*({\bf{R}})~d{\bf{R}}
\end{eqnarray}
and $ \delta_{{\boldsymbol{\lambda}},\boldsymbol{\lambda'}}$ 
is the Kronecker delta symbol.
The determination of 
$I_{{\boldsymbol{\lambda}},\boldsymbol{\lambda'}}^{\rm{3D}}(\epsilon_{\boldsymbol{\lambda'}})$
and $\mathcal{F}^{\rm{3D}}(\epsilon_{\boldsymbol{\lambda'}},\eta)$
for $\eta > 1$ and $\eta < 1$ is discussed in 
Secs.~{\ref{sec_cigar}} and~{\ref{sec_pancake}}, respectively.

Equation~(\ref{eq_eigen2}) can be interpreted as 
a matrix equation with eigenvalues $a_z/a^{\rm{3D}}$ 
and eigenvectors $f_{\boldsymbol{\lambda}}$~\cite{kestner, drummond_prl}.
In practice, we first calculate the
 matrix elements
$I_{{\boldsymbol{\lambda}},\boldsymbol{\lambda'}}^{\rm{3D}}(\epsilon_{\boldsymbol{\lambda'}})$
 in Eq.~(\ref{eq_eigen2}) for a given 
three-body energy $E_{\rm{3b}}$ and obtain 
the corresponding scattering lengths for this energy by
diagonalizing the matrix with elements
$I_{{\boldsymbol{\lambda}},\boldsymbol{\lambda'}}^{\rm{3D}}(\epsilon_{\boldsymbol{\lambda'}})-
\mathcal{F}^{\rm{3D}}(\epsilon_{\boldsymbol{\lambda}},\eta) \delta_{{\boldsymbol{\lambda}},\boldsymbol{\lambda'}}$.
This step is repeated for several three-body energies.
Lastly, we invert $a^{\rm{3D}}(E_{\rm{3b}})$ to get $E_{\rm{3b}}(a^{\rm{3D}})$, i.e.,
to get the three-body energies as a function of the $s$-wave
scattering length.

Equation~(\ref{eq_eigen2}) has a simple physical interpretation.
If the interaction between particles 2 and 3 is turned off, the matrix
$I_{{\boldsymbol{\lambda}},\boldsymbol{\lambda'}}^{\rm{3D}}(\epsilon_{\boldsymbol{\lambda'}})$
vanishes and the solution reduces to that of an interacting pair
(particles 1 and 3) and a non-interacting spectator particle (particle 2).
The relative 
energy $(\epsilon_{\boldsymbol{\lambda'}}+\eta+1/2)E_z$
of the pair is determined by solving the relative two-body eigenequation
$\mathcal{F}^{\rm{3D}}(\epsilon_{\boldsymbol{\lambda'}},\eta)=-a_z/a^{\rm{3D}}$.
The matrix
$I_{{\boldsymbol{\lambda}},\boldsymbol{\lambda'}}^{\rm{3D}}(\epsilon_{\boldsymbol{\lambda'}})$
thus arises from the fact that particle 3 not only 
interacts with particle 1 but also with particle 2.
Correspondingly, the terms in Eq.~(\ref{eq_eigen2}) that contain 
$I_{{\boldsymbol{\lambda}},\boldsymbol{\lambda'}}^{\rm{3D}}(\epsilon_{\boldsymbol{\lambda'}})$
can be interpreted as exchange terms that arise from
exchanging particles 1 and 2~\cite{drummond_prl}.

For $\eta = 1$, the function 
$\mathcal{F}^{\rm{3D}}(\epsilon_{\boldsymbol{\lambda}},\eta)$
is given in Table~\ref{tab_two_scatt} and the evaluation of
$I_{{\boldsymbol{\lambda}},\boldsymbol{\lambda'}}^{\rm{3D}}(\epsilon_{\boldsymbol{\lambda'}})$
has been discussed in detail in Ref.~\cite{drummond_pra}.
The $\eta \neq 1$ cases are discussed in Secs.~{\ref{sec_cigar}} and ~{\ref{sec_pancake}}.
\begin{table}
\caption{Two-body properties in three dimensions 
($s$-wave channel), two dimensions ($m=0$ channel), 
and one dimension (even parity channel). 
$\psi({\bf{q}})$ denotes the relative two-body wave function 
and $\mu$ the two-body reduced mass. ${\bf{q}}$ stands for  
${\bf{r}}$, ${\boldsymbol{\rho}}$, and $z$ in three, two, 
and one dimensions, respectively.
For 3D, 2D and 1D, we have $\epsilon=E_{\rm{2b}}/E_z-\eta-1/2$,
$\epsilon=E_{\rm{2b}}/E_{\rho}-1$ and $\epsilon=E_{\rm{2b}}/E_z-1/2$, respectively,
where $E_{\rm{2b}}$ denotes the relative two-body energy.
$\psi_g$ denotes the digamma function and $\gamma$ the Euler constant,
$\gamma \approx 0.577$.}
\begin{ruledtabular}
\begin{tabular}{llll}
 & 
$\rm{3D}$ & 
$\rm{2D}$ & 
$\rm{1D}$\\
\hline
$V_{\rm{ps}}$ & 
$g^{\rm{3D}}\delta^{(3)}({\bf{r}})\frac{\partial}{\partial r}r$ & 
$g^{\rm{2D}}\delta^{(2)}({\boldsymbol{\rho}})\frac{\partial}{\partial \rho}\rho$ & 
$g^{\rm{1D}}\delta(z)$ \\
\hline
$g$ & 
$2\pi \frac{\hbar^2}{\mu}a^{\rm{3D}}$ & 
$\pi \frac{\hbar^2}{\mu} \left[\ln(\frac{\rho}{a^{\rm{2D}}})+1\right]^{-1}$ & 
$ -\frac{\hbar^2}{\mu}\frac{1}{a^{\rm{1D}}}$ \\
\hline
$ \left.\psi({\bf{q}}) \right|_{|{\bf{q}}| \to 0} $ & 
$\propto \left(\frac{1}{r}-\frac{1}{a^{\rm{3D}}}\right)$ & 
$\propto \left[\ln(a^{\rm{2D}})-\ln(\rho)\right]$ & 
$\propto (z- a^{\rm{1D}})$ \\
\hline
Bethe-Peierls B.C. & 
$\left . \frac{\partial (r \psi)}{\partial r} \right|_{r \to 0} = \frac{-1}{a^{\rm{3D}}}\left . (r \psi) \right|_{r \to 0}$ & 
$\left . \rho \frac{\partial \psi}{\partial \rho} \right|_{\rho \to 0} = \left . \frac{\psi}{\ln\left(\rho/a^{\rm{2D}}\right)} \right|_{\rho \to 0}$ & 
$\left . \frac{d \psi}{d z} \right|_{z \to 0} = \frac{-1}{a^{\rm{1D}}}\left . \psi \right|_{z \to 0}$ \\
\hline
$\mathcal{F}$ & 
$\mathcal{F}^{\rm{3D}}(\epsilon,\eta)= 2 \pi E_z a_z^3$&
$\mathcal{F}^{\rm{2D}}(\epsilon)= \pi E_{\rho} a_{\rho}^2$&
$\mathcal{F}^{\rm{1D}}(\epsilon)= E_z a_z$
\\ & $ \left\{ \frac{\partial}{\partial r}\left[r\mathcal{G}^{\rm{3D}}\left([\epsilon+\eta+\frac{1}{2}]E_z;{\bf{r}};{\bf{0}}\right)\right]\right\}_{r \rightarrow 0}$&
$\left\{  \mathcal{G}^{\rm{2D}}\left([\epsilon+1]E_\rho;{{\rho}};{{0}}\right) + \ln(\rho/a_{\rho})\right\}_{\rho \rightarrow 0}$& 
$\left\{ \mathcal{G}^{\rm{1D}}\left([\epsilon+\frac{1}{2}]E_z;z;0\right)\right\}_{z \rightarrow 0}$
\\
\hline
&$\mathcal{F}^{\rm{3D}}(\epsilon,1)=\frac{-2\Gamma(-\epsilon/2)}{\Gamma(-\epsilon/2-1/2)}$ & 
$\mathcal{F}^{\rm{2D}}(\epsilon)=-\frac{1}{2}\psi_g(-\epsilon/2) - \gamma$ & 
$\mathcal{F}^{\rm{1D}}(\epsilon)=\frac{\Gamma(-\epsilon/2)}{2 \Gamma(-\epsilon/2+ 1/2)}$ \\
\hline
two-body energy &
$\mathcal{F}^{\rm{3D}}(\epsilon, \eta) = -a_z/a^{\rm{3D}}$ & 
$\mathcal{F}^{\rm{2D}}(\epsilon) = \ln(a^{\rm{2D}}/a_{\rho})$ & 
$\mathcal{F}^{\rm{1D}}(\epsilon) = a^{\rm{1D}}/a_z$ \\
\end{tabular}
\end{ruledtabular}
\label{tab_two_scatt}
\end{table}
For a spherically symmetic system with $\eta = 1$,
the total relative angular momentum quantum number $L$,
the corresponding projection quantum number $M$ 
and the parity $\Pi$ are
good quantum numbers, and the eigenvalue equation can be solved 
for each $L$ and $M$ combination separately
using spherical coordinates~\cite{kestner}. 
For a fixed $L$ and $M$,
$\boldsymbol{\lambda}=(n,l,m)$ and
$\boldsymbol{\lambda'}=(n',l',m')$ in Eq.~(\ref{eq_eigen2})
are constrained by $l=l'=L$ and $m=m'=M$.
The parity of the three-body system is given by 
$\Pi=(-1)^L$.

We emphasize that the outlined formalism makes no approximations, 
i.e., Eq.~(\ref{eq_eigen2}) with 
$I_{{\boldsymbol{\lambda}},\boldsymbol{\lambda'}}^{\rm{3D}}
\left(\epsilon_{\boldsymbol{\lambda'}}\right)$
given by Eq.~(\ref{eq_Ilambda})
describes all eigenstates of 
$H_{\rm{rel}}$ 
[see Eq.~(\ref{eq_rel_Hamiltonian})]
that are affected by the interactions.
In particular, all ``channel couplings'' are accounted for.
In practice, the construction of the matrix
$I_{{\boldsymbol{\lambda}},\boldsymbol{\lambda'}}^{\rm{3D}}(\epsilon_{\boldsymbol{\lambda'}})$
requires one to choose a maximum for ${\boldsymbol{\lambda}}$ 
and $\boldsymbol{\lambda'}$, or alternatively,
a cutoff for the single-particle energy 
$E_{\boldsymbol{\lambda}}$.
As has been shown in Ref.~\cite{drummond_pra},
good convergence is 
achieved for a relatively small number of ``basis functions''
for $\eta=1$. As we show below, good convergence is also
obtained for anisotropic confinement geometries.

The formalism outlined can also be applied to strictly 
one-dimensional and strictly two-dimensional systems.
Table~\ref{tab_two_scatt} defines the one-dimensional 
and two-dimensional pseudopotentials as well as a 
number of key properties of the corresponding relative two-body system.
Making the appropriate changes in the outlined derivation
and using the properties listed in Table~\ref{tab_two_scatt},
we find for strictly one-dimensional systems
\begin{eqnarray}
\label{eq_eigen_1d}
\sum_{n_z'=0}^{\infty} 
\left[
I_{n_z,n_z'}^{\rm{1D}}\left(\epsilon_{n_z'}\right) 
- \mathcal{F}^{\rm{1D}}\left(\epsilon_{n_z}\right) 
\delta_{n_z,n_z'} 
\right]
f_{n_z'}=
-\frac{a^{\rm{1D}}}{a_z}f_{n_z},
\end{eqnarray}
where $\mathcal{F}^{\rm{1D}}$ is defined in Table~\ref{tab_two_scatt},
 \begin{eqnarray}
\label{eq_Inz_1d}
I_{n_z,n_z'}^{\rm{1D}}\left(\epsilon_{n_z'}\right)=
E_z a_z\int_{-\infty}^{\infty}
\mathcal{G}^{\rm{1D}}\left([\epsilon_{n_z'}+1/2]E_z;
\frac{\sqrt{3}}{2}z;0\right)
\varphi_{n_z'}\left(-\frac{z}{2}\right)\varphi_{n_z}^*(z) dz,
\end{eqnarray}
and $E_{\rm{3b}} - E_{n_z'} = (\epsilon_{n_z'} + 1/2)E_z$.
Here, $E_{n_z}$ denotes the single-particle energy of the 
one-dimensional system, $E_{n_z}=(n_z+1/2) E_z$,
and $\mathcal{G}^{\rm{1D}}\left(E;z;z'\right)$ the one-dimensional 
even parity single-particle Green's function,
\begin{eqnarray}
\label{eq_one_green_1d}
\mathcal{G}^{\rm{1D}}\left(E;z;z'\right) = 
\sum_{n_z'=0}^{\infty}\frac{\varphi_{2n_z'}^*(z')
\varphi_{2n_z'}(z)
}{E_{2n_z'}-E}.
\end {eqnarray}
For $z'=0$, the single-particle Green's function is given by
\begin{align}
\label{eq_2-body_1d}
\mathcal{G}^{\rm{1D}}\left(E;z;0\right) 
= \frac{1}{2\sqrt{\pi} E_z a_z}
\exp\left(-\frac{z^2}{2 a_z^2}\right)
\Gamma\left(-\frac{E/E_z-1/2}{2}\right) 
U\left(-\frac{E/E_z-1/2}{2},\frac{1}{2},\frac{z^2}{a_z^2}\right),
\end{align}
where $\Gamma(x)$ is the Gamma function and $U(a,b,z)$ the confluent
hypergeometric function.
The strictly one-dimensional relative three-body wave function
$\Psi$ is characterized by the parity $\Pi_z$.
For even parity states, i.e., for states with $\Pi_z=1$,
$n_z$ and $n_z'$ in Eq.~(\ref{eq_eigen_1d}) have to be even.
For odd parity states, i.e., for states with $\Pi_z=-1$,
$n_z$ and $n_z'$ have to be odd.

Similarly, for strictly two-dimensional systems, expressed in
units of $E_{\rho}$ and $a_{\rho}$ [$E_{\rho}=\hbar \omega_{\rho}$
and $a_{\rho}=\sqrt{\hbar/(\mu \omega_{\rho})}$], 
we find, in agreement with Ref.~\cite{drummond_2d},
\begin{eqnarray}
\label{eq_eigen_2d}
\sum_{n_{\rho}'=0}^{\infty} 
\left[
I_{n_{\rho},n_{\rho}',m}^{\rm{2D}}(\epsilon_{n_{\rho}',m}) - 
\mathcal{F}^{\rm{2D}}(\epsilon_{n_{\rho},m}) 
\delta_{n_{\rho},n_{\rho}'}
\right]
f_{n_{\rho}',m}=
\ln\left(\frac{a_{\rho}}{a^{\rm{2D}}}\right)f_{n_{\rho},m},
\end{eqnarray}
where $\mathcal{F}^{\rm{2D}}$ is defined in Table~\ref{tab_two_scatt},
 \begin{eqnarray}
\label{eq_Inrho_2d}
I_{n_{\rho},n_{\rho}',m}^{\rm{2D}}(\epsilon_{n_{\rho}',m})= 
(-1)^m ~\pi 
E_{\rho} a_{\rho}^2
\int_0^{\infty}
\mathcal{G}^{\rm{2D}}
\left([\epsilon_{n_{\rho}',m}+1]E_{\rho};\frac{\sqrt{3}}{2}\rho;0\right)
R_{n_{\rho}',m}\left(\frac{\rho}{2}\right)
R_{n_{\rho},m}\left(\rho\right) \rho d \rho,
\end{eqnarray}
and $E_{\rm{3b}}-E_{n_{\rho}',m}=(\epsilon_{n_{\rho}',m}+1) E_{\rho}$.
Here, $E_{n_{\rho},m}$ denotes the single-particle energy of the 
two-dimensional system, $E_{n_{\rho},m}= (2 n_{\rho} + |m| +1)E_{\rho}$.
The two-dimensional single-particle Green's function
$\mathcal{G}^{\rm{2D}}\left(E;{\rho}; {\rho}'\right)$
is defined analogously to the three- and one-dimensional counterparts 
[see Eqs.~(\ref{eq_one_green}) and~(\ref{eq_one_green_1d})].
For $\rho'=0$ and states affected 
by the zero-range $s$-wave interactions~\cite{calarco-r,calarco}, 
one finds
\begin{align}
\label{eq_2-body_2d}
\mathcal{G}^{\rm{2D}}(E;{\rho};{0}) = 
\frac{1}{2\pi E_{\rho} a_{\rho}^2}
\exp\left(-\frac{\rho^2}{2 a_{\rho}^2}\right)
\Gamma\left(-\frac{E/E_{\rho}-1}{2}\right) 
U\left(-\frac{E/E_{\rho}-1}{2},1,\frac{\rho^2}{a_{\rho}^2}\right).
\end{align}
The strictly two-dimensional relative three-body wave function
is characterized by the projection quantum number $M$ 
and the parity $\Pi_{\boldsymbol{\rho}}$, 
$\Pi_{\boldsymbol{\rho}}=(-1)^{M}$.
For a fixed $M$, $m$ in Eq.~(\ref{eq_eigen_2d})
is constrained to the value $m=M$.
The next two sections analyze, utilizing our results for strictly
one- and two-dimensional systems, Eq.~(\ref{eq_eigen2}) 
for cigar- and pancake-shaped traps.

\section{Cigar-shaped trap}
\label{sec_cigar}
To apply the formalism reviewed in 
Sec.~{\ref{sec_formalsolution}} to axially symmetric traps,
we need the 
explicit forms of the functions 
$\mathcal{G}^{\rm{3D}}\left([\epsilon_{\boldsymbol{\lambda}}+\eta+1/2]E_z;
{\bf{r}};{\bf{0}}\right)$
and
$\mathcal{F}^{\rm{3D}}(\epsilon_{\boldsymbol{\lambda}},\eta)$,
that is, the relative solutions to the trapped two-body system.
For cigar-shaped traps ($\eta > 1$), it is convenient 
to write
$\mathcal{G}^{\rm{3D}}$  as~\cite{calarco-r,calarco}
\begin{align}
\label{eq_G_cigar}
\mathcal{G}^{\rm{3D}}\left([\epsilon_{\boldsymbol{\lambda}}+\eta+1/2]E_z;
{\bf{r}};{\bf{0}}\right)=
\frac{\eta}{\pi a_z^2}\exp\left(-\frac{ \eta \rho^2}{2 a_z^2}\right)
\sum_{j=0}^{\infty} L_j\left( \eta \rho^2 / a_z^2\right) 
\mathcal{G}^{\rm{1D}}\left([\epsilon_{\boldsymbol\lambda}-2 \eta j+ 
1/2]E_z ;z;0\right),
\end{align}
where $\mathcal{G}^{\rm{1D}}\left(E;z;0\right)$
is defined in Eq.~(\ref{eq_2-body_1d}).
Using Eq.~(\ref{eq_G_cigar}) in Eq.~(\ref{eq_Ilambda}), we obtain
\begin{eqnarray}
\label{eq_Ilambda_cigar}
I_{{\boldsymbol{\lambda}},\boldsymbol{\lambda'}}^{\rm{3D}}
(\epsilon_{\boldsymbol{\lambda'}}) =
\sqrt{2\eta}(-1)^m\delta_{m,m'}\times \nonumber \\
\lim_{j_{\rm{max}} \to \infty}\sum_{j=0}^{j_{\rm{max}}}I_{n_z,n_z'}^{\rm{c}}(\epsilon_{\boldsymbol{\lambda'}},j) I_{n_{\rho},n_{\rho}',m}^{\rm{c}}(j),
\end{eqnarray}
where
\begin{align}
\label{eq_Iz_cigar}
 I_{n_z,n_z'}^{\rm{c}}(\epsilon_{\boldsymbol{\lambda'}},j)=
E_z a_z
\int_{-\infty}^{\infty}
\mathcal{G}^{\rm{1D}}\left([\epsilon_{\boldsymbol{\lambda'}}-2 \eta j+
1/2]E_z;\frac{\sqrt{3}}{2}z; 0\right)
 \varphi_{n_z'}\left(-\frac{z}{2}\right)\varphi_{n_z}(z)~dz
\end{align}
and
\begin{align}
\label{eq_Irho_cigar}
& I_{n_{\rho},n_{\rho}',m}^{\rm{c}}(j)=\nonumber \\
&a_z\int_0^{\infty} R_{j,0}\left(\frac{\sqrt{3}}{2}\rho\right) R_{n_{\rho}',m}\left(\frac{\rho}{2}\right) R_{n_{\rho},m}(\rho)~\rho d\rho.
\end{align}
The evaluation of the integrals 
$I_{n_z,n_z'}^{\rm{c}}(\epsilon_{\boldsymbol{\lambda'}},j)$ and 
$I_{n_{\rho},n_{\rho}',m}^{\rm{c}}(j)$ is discussed in
Appendix~\ref{appendix}.
The superscript ``c'' indicates that the integrals
apply to cigar-shaped systems; for pancake-shaped systems
(see Sec.~\ref{sec_pancake}), we introduce the integrals 
$I_{n_{\rho},n_{\rho}',m}^{\rm{p}}(\epsilon_{\boldsymbol{\lambda'}},j)$ and 
$I_{n_z,n_z'}^{\rm{p}}(j)$ 
instead.

Although it is possible to calculate
$\mathcal{F}^{\rm{3D}}(\epsilon_{\boldsymbol{\lambda}},\eta)$
numerically for any trap aspect ratio $\eta$, we
restrict ourselves to integer aspect ratios for simplicity.
For traps with integer aspect ratio, an exact analytical expression for 
$\mathcal{F}^{\rm{3D}}(\epsilon_{\boldsymbol{\lambda}},\eta)$ is
known~\cite{calarco-r,calarco},
\begin{eqnarray}
\label{eq_f26_cigar}
\mathcal{F}^{\rm{3D}} \left( \epsilon_{\boldsymbol{\lambda}},\eta \right)=
-2 \frac
{\Gamma\left(-\frac{\epsilon_{\boldsymbol{\lambda}}}{2}\right)}
{\Gamma\left(-\frac{1}{2}-\frac{\epsilon_{\boldsymbol{\lambda}}}{2}\right)} 
+ \frac
{\Gamma\left(-\frac{\epsilon_{\boldsymbol{\lambda}}}{2}\right)}
{\Gamma\left(\frac{1}{2}-\frac{\epsilon_{\boldsymbol{\lambda}}}{2}\right)} 
\times \nonumber \\
\sum _{k=1}^{\eta -1} {}_2F_1 \left(1,-\frac{\epsilon_{\boldsymbol{\lambda}}}{2};\frac{1}{2}-\frac{\epsilon_{\boldsymbol{\lambda}}}{2};\exp
\left(\frac{2 \pi \imath k}{\eta} \right) \right),
\end{eqnarray}
where
$ {}_2F_1(a,b;c;z)$ is the hypergeometric function~\cite{abramowitz}.
Knowing $I_{\boldsymbol{\lambda},\boldsymbol{\lambda'}}^{\rm{3D}}
(\epsilon_{\boldsymbol{\lambda'}})$
and $\mathcal{F}^{\rm{3D}}(\epsilon_{\boldsymbol{\lambda'}},\eta)$,
Eq.~(\ref{eq_eigen2}) can be diagonalized separately for each 
$(\Pi_z, M,\Pi_{\boldsymbol{\rho}})$ combination. 
We recall from Sec.~\ref{sec_formalsolution} that 
$\boldsymbol{\lambda}=(n_z,n_{\rho},m)$ and  
$\boldsymbol{\lambda}'=(n_z',n_{\rho}',m')$.
The $m$ and $m'$ values are constrained by
$m=m'=M$.
Moreover, for $\Pi_z=+1$ and $\Pi_z=-1$, we have
 $n_z=n_z'=even$ and $n_z=n_z'=odd$, respectively.

\begin{figure}
\vspace*{+.9cm}
\includegraphics[angle=0,width=70mm]{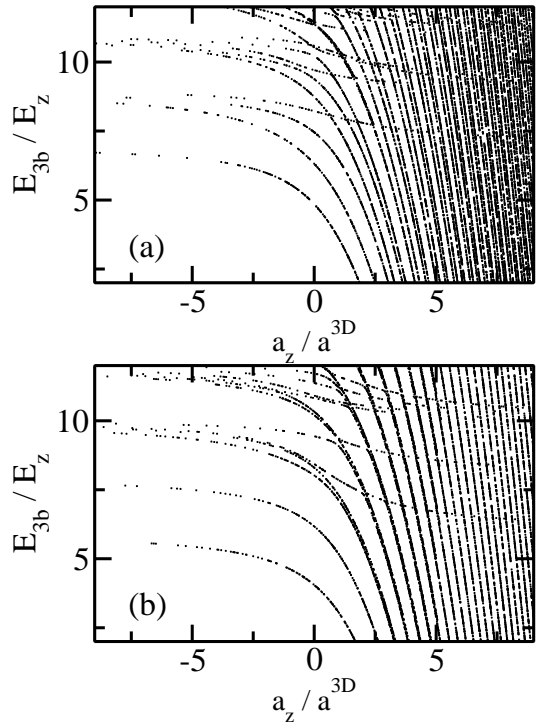}
\vspace*{0.3cm}
\caption{Relative three-body energies $E_{\rm{3b}}/E_z$ as a 
function of the 
inverse scattering length $a_z/a^{\rm{3D}}$ for a cigar-shaped trap 
with aspect ratio $\eta = 2$ and 
(a) $M=0$ and $\Pi_z = +1$, and (b) $M=0$ and $\Pi_z = -1$.
}
\label{fig_cigar}
\end{figure}

Figure~\ref{fig_cigar}
shows the three-body relative energies $E_{\rm{3b}}/E_z$
for $\eta=2$ for states with (a) $M=0$ and $\Pi_z = +1$
and (b) $M=0$ and $\Pi_z = -1$ 
as a function of the inverse scattering length
$a_z/a^{\rm{3D}}$.
The
non-interacting limit is approached when $(a^{\rm{3D}})^{-1} \to \pm \infty$,
and the infinitely strongly-interacting regime for  $(a^{\rm{3D}})^{-1}=0$
(center of the figure).
For 
each fixed projection quantum number $M$, 
we include around $840$ basis functions.
This corresponds to a cutoff of around $(82+2M)E_z$
for the single-particle energy $E_{\boldsymbol{\lambda}}$.
We find that $j_{\rm{max}} \gtrsim 30$ yields converged values for  
$I_{{\boldsymbol{\lambda}},\boldsymbol{\lambda'}}^{\rm{3D}}(\epsilon_{\boldsymbol{\lambda'}})$,
Eq.~(\ref{eq_Ilambda_cigar}).
For small $|a^{\rm{3D}}/a_z|$ ($a^{\rm{3D}}$ positive and negative), 
our eigenenergies agree with those obtained within first-order
perturbation theory.
Our analysis shows that the energy of the ground state
at unitarity has a relative 
error of the order of $10^{-5}$.
The accuracy decreases with increasing energy. 
For example, for energies around $20 E_z$, 
the relative accuracy at unitarity is of the order of $10^{-4}$.

The eigenstates fall into one of two categories:
atom-dimer states and atom-atom-atom states.
The eigenenergies associated with the former are negative for 
large positive $a_z/a^{\rm{3D}}$ while those associated with the 
latter remain positive for large positive $a_z/a^{\rm{3D}}$.
The energy spectra shown in Fig.~\ref{fig_cigar}
exhibit sequences of avoided crossings.
To resolve these crossings, a fairly fine mesh in the three-body energy 
is needed.
In the $(a^{\rm{3D}})^{-1} \to - \infty$ limit, the lowest
$M=0$ state has negative parity in $z$, i.e., $\Pi_z = -1$.
This is a direct consequence of the fact that the two identical fermions
cannot occupy the same single particle state.
In the $(a^{\rm{3D}})^{-1} \to + \infty$ limit, in contrast, the lowest
$M=0$ state has positive parity in $z$, i.e., $\Pi_z = +1$.
This is a direct consequence of the fact that the system
consists, effectively, of a dimer and an atom.

\begin{figure}
\vspace*{+.9cm}
\includegraphics[angle=0,width=70mm]{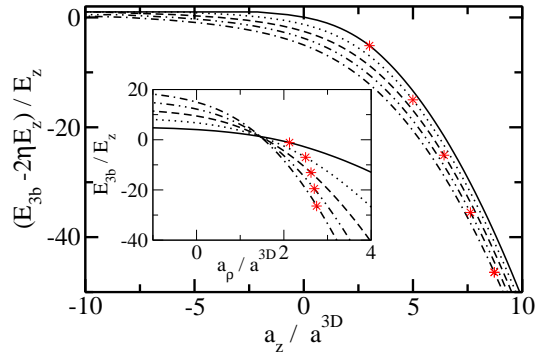}
\vspace*{0.5cm}
\caption{(Color online)~``Crossover curve'' of the 
three-body system with $M=0$, shifted by $2 E_{\rho} = 2 \eta E_z$, 
for cigar-shaped traps with $\eta = 2$ (solid line), $4$ (dotted line), 
$6$ (dashed line), $8$ (dash-dot-dotted line), 
and $10$ (dash-dotted line) as a function of $a_z/a^{\rm{3D}}$.
The scattering lengths at which the parity of the corresponding 
eigenstate changes from 
$\Pi_z = -1$ (``left side of the graph'') to 
$\Pi_z = +1$ (``right side of the graph'') are marked by asterisks.
At these points, the derivative of the crossover curve
is discontinuous;
the discontinuities are not visible on the scale shown.
The inset shows the (unshifted)
crossover curve as a function of $a_{\rho}/a^{\rm{3D}}$.}
\label{fig_cigar_gr_st}
\end{figure}

The main part of Fig.~\ref{fig_cigar_gr_st}
shows the relative energy of the energetically 
lowest-lying state, the so-called crossover curve, 
of the three-body system with $M=0$ for various aspect ratios
of the trap ($\eta = 2,\cdots,10$)
as a function of the inverse scattering length $a_z/a^{\rm{3D}}$.
For comparative purposes, we subtract the ground state
energy of $2\eta E_z$ of the strictly two-dimensional non-interacting system,
that is, the energy that the system would have in the
$\rho$-direction if the dynamics in the tight confinement 
direction were frozen,
from the full three-dimensional energy.
In Fig.~\ref{fig_cigar_gr_st}, asterisks mark the scattering lengths 
at which the eigenstate associated with the crossover curve changes from 
$\Pi_z = -1$ to $\Pi_z = +1$.
With increasing $\eta$, the parity change occurs at larger
$a_z / a^{\rm{3D}}$ (that is, smaller $a^{\rm{3D}}/a_z$).
The inset of Fig.~\ref{fig_cigar_gr_st} replots the crossover
curves as a function of $a_{\rho}/a^{\rm{3D}}$. 

\begin{figure}
\vspace*{+1.9cm}
\includegraphics[angle=0,width=70mm]{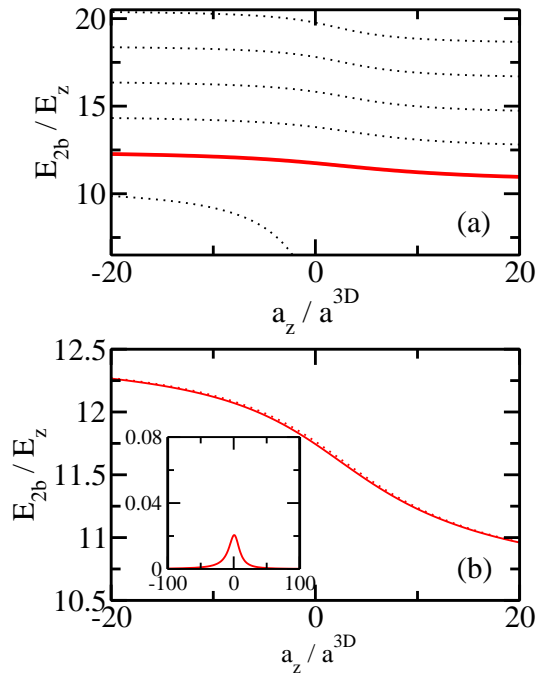}
\vspace*{0.5cm}
\caption{(Color online)~(a)~Relative two-body energies $E_{\rm{2b}}/E_z$ 
as a function of the 
inverse scattering length $a_z/a^{\rm{3D}}$ for a cigar-shaped trap 
with aspect ratio $\eta = 10$, $M=0$ and $\Pi_z=+1$.
(b)~The solid curve from panel (a) is replotted and compared with the
energy obtained by solving the strictly one-dimensional 
eigenequation with renormalized 
one-dimensional scattering length $a_{\rm{ren}}^{\rm{1D}}$
(dotted line; 
the energy $E_{\rho}=\eta E_z$ has been added to allow for a comparison with 
the full three-dimensional energy).
The inset shows the difference between the dotted and solid lines as 
a function of the inverse scattering length $a_z/a^{\rm{3D}}$.
The scale of the $y$-axis is identical to that of the 
inset of Fig.~\ref{fig_cigar_quasi_3-body}(b).}
\label{fig_cigar_quasi_2-body}
\end{figure}

We now discuss the large $\eta$ limit in more detail.
Using the limiting behavior of 
$\mathcal{F}^{\rm{3D}}(\epsilon, \eta)$
for $\eta\gg 1$ and 
$\eta\gg |\epsilon|$~\cite{calarco-r,calarco},
\begin{align}
\label{eq_f_large_eta}
\left. 
\mathcal{F}^{\rm{3D}}
(\epsilon, \eta)\right|_{\eta\gg 1}
\approx 
2\eta  \mathcal{F}^{\rm{1D}}(\epsilon)
+\sqrt{\eta} \zeta(1/2),
\end{align}
the two-body eigenequation for the relative energy 
becomes~\cite{calarco-r,calarco}
\begin{align}
\label{eq_two-b_1d}
\mathcal{F}^{\rm{1D}}(\epsilon)=\frac{a_{\rm{ren}}^{\rm{1D}}}{a_z},
\end{align}
where the renormalized one-dimensional scattering length
$a_{\rm{ren}}^{\rm{1D}}$ is given by~\cite{olshanii1,olshanii2}
\begin{align}
\label{eq_olshanii}
\frac{a_{\rm{ren}}^{\rm{1D}}}{a_z}=\frac{1}{\sqrt{\eta}} \left[
-\frac{a_{\rho}}{2 a^{\rm{3D}}}-\frac{\zeta(1/2)}{2}
\right].
\end{align}
Figure~\ref{fig_cigar_quasi_2-body}(a)
shows the relative two-body energies for a system with 
$\eta = 10$, $M=0$ and $\Pi_z=+1$ obtained by solving
the eigenequation
$\mathcal{F}^{\rm{3D}}(\epsilon, \eta=10)=-a_z/a^{\rm{3D}}$
[see Eq.~(\ref{eq_f26_cigar}) for $\mathcal{F}^{\rm{3D}}(\epsilon, \eta)$].
Figure~\ref{fig_cigar_quasi_2-body}(b)
compares the full three-dimensional energy (solid line)
with the energy obtained by solving the strictly 
one-dimensional eigenequation, Eq.~(\ref{eq_two-b_1d}),
with renormalized one-dimensional scattering length $a_{\rm{ren}}^{\rm{1D}}$
(dotted line).
To facilitate the comparison, we add the energy of the tight 
confinement direction to the energy of the one-dimensional system.
The agreement is quite good for all scattering lengths.
The inset of Fig.~\ref{fig_cigar_quasi_2-body}(b)
shows the difference between the strictly one-dimensional energy
and the full three-dimensional energy as a function of $a_z/a^{\rm{3D}}$.
The maximum deviation occurs around unitarity and is of the 
order of $0.2 \%$.

Next, we discuss the behavior of the three-body system
in the large $\eta$ limit.
If we use Eqs.~(\ref{eq_f_large_eta}) and~(\ref{eq_olshanii}) 
in Eq.~(\ref{eq_eigen2}),
we find 
\begin{align}
\label{eq_eigen_large_eta}
\sum_{\boldsymbol{\lambda'}}\left[
 \frac{1}{2\eta} I^{\rm{3D}}_{\boldsymbol{\lambda},\boldsymbol{\lambda'}}(\epsilon_{\boldsymbol{\lambda}'})-
\mathcal{F}^{\rm{1D}}(\epsilon_{\boldsymbol{\lambda}})
\delta_{\boldsymbol{\lambda},\boldsymbol{\lambda}'}
 \right]f_{\boldsymbol{\lambda}'}
= - \frac{a_{\rm{ren}}^{\rm{1D}}}{a_z}f_{\boldsymbol{\lambda}}.
\end{align}
A straightforward analysis shows that Eq.~(\ref{eq_eigen_large_eta})
reduces to its strictly one-dimensional analog if
{\em{(i)}} the sum over
$\boldsymbol{\lambda}'$ is restricted to a sum over $n_z'$ 
[$\boldsymbol{\lambda}'=(n_z',0,0)$];
{\em{(ii)}} the index $\boldsymbol{\lambda}$ is restricted to 
$\boldsymbol{\lambda}=(n_z,0,0)$;
{\em{(iii)}} the energy $E_{\rm{3b}}$ is replaced by $E_{\rm{3b}} -2\eta E_z$;
and
{\em{(iv)}} $j_{\rm{max}}$ in Eq.~(\ref{eq_Ilambda_cigar}) is set to zero.
Under these assumptions,  Eq.~(\ref{eq_eigen_large_eta})
reduces to Eq.~(\ref{eq_eigen_1d}) with $a^{\rm{1D}}$ 
replaced by $a_{\rm{ren}}^{\rm{1D}}$.
We emphasize that the assumption
$\eta\gg |\epsilon_{\boldsymbol{\lambda}}|$
[see discussion around Eq.~(\ref{eq_f_large_eta})]
is not valid when two 
atoms form a tight molecule. In this limit, the 
three-dimensional $s$-wave scattering length, 
or the size of the dimer, is smaller 
than the harmonic oscillator length in the transverse direction,
which implies that the strictly one-dimensional
description is not valid.

\begin{figure}
\vspace*{+.9cm}
\includegraphics[angle=0,width=70mm]{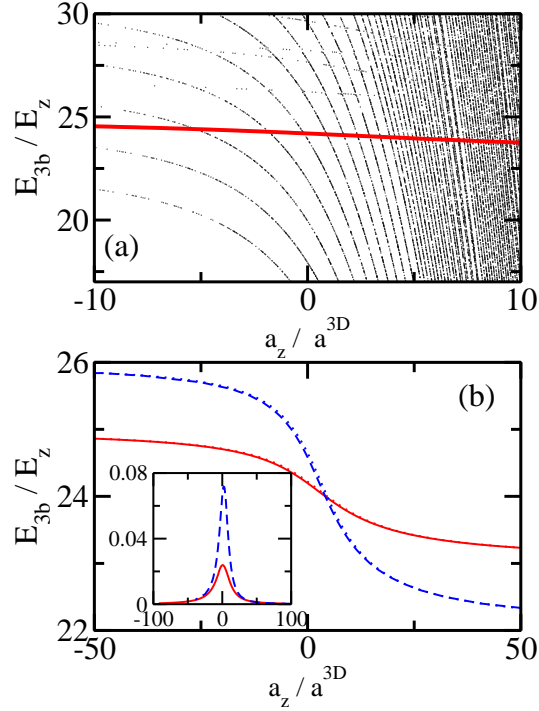}
\vspace*{0.4cm}
\caption{(Color online)~(a)~Relative three-body 
energies $E_{\rm{3b}}/E_z$ as a function of the 
inverse scattering length $a_z/a^{\rm{3D}}$ for a cigar-shaped
trap with aspect ratio $\eta = 10$, $M=0$ and $\Pi_z=+1$.
(b)~The solid curve from panel (a) is replotted and compared with the
energy obtained by solving the strictly one-dimensional 
eigenequation with renormalized 
one-dimensional scattering length $a_{\rm{ren}}^{\rm{1D}}$
(the energy of $2 E_{\rho}=2\eta E_z$ 
has been added to allow for a comparison with 
the full three-dimensional energy).
For comparison, the dashed line shows one of the three-dimensional 
energy curves for
$M=0$ and $\Pi_z=-1$ [not shown in panel~(a)];
the corresponding strictly one-dimensional energy is shown by a dotted line.
The difference between the full three-dimensional and
strictly one-dimensional descriptions is hardly visible on the scale shown.
Solid and dashed lines
in the inset show the differences between 
the strictly 
one-dimensional energies 
[dotted lines in panel~(b)] and 
the full three-dimensional energies 
[solid and dashed lines in panel~(b)]
as 
a function of the inverse scattering length $a_z/a^{\rm{3D}}$
for $\Pi_z=+1$ and $\Pi_z=-1$, respectively.}
\label{fig_cigar_quasi_3-body}
\end{figure}

Figure~\ref{fig_cigar_quasi_3-body}(a)
shows the relative three-body energies for states with 
$M=0$ and $\Pi_z=+1$
as a function of the inverse scattering length $a_z/a^{\rm{3D}}$
for a cigar-shaped trap with $\eta=10$.
Figure~\ref{fig_cigar_quasi_3-body}(b)
compares the energy of the energetically lowest-lying
three-atom state with $\Pi_z=+1$ (solid line)
[see thick solid line in Fig.~\ref{fig_cigar_quasi_3-body}(a)]
with the corresponding state obtained by solving
the strictly one-dimensional equation with renormalized 
one-dimensional scattering length $a_{\rm{ren}}^{\rm{1D}}$.
We also include the energy of one of the eigenstates with 
$M=0$ and $\Pi_z= -1$ (dashed line).
The inset shows the difference between the energy obtained 
within the strictly one-dimensional and the full 
three-dimensional frameworks.
The maximum of the deviation occurs near unitarity.
The agreement between the full three-dimensional and the 
strictly one-dimensional descriptions is good. 
Importantly, the deviations for the three-body system
with $M=0$ and $\Pi_z=+1$
[solid line in the inset of Fig.~\ref{fig_cigar_quasi_3-body}(b)]
are only slightly larger than those for the two-body system
[inset of Fig.~\ref{fig_cigar_quasi_2-body}(b)],
suggesting that the presence of the third atom does not, in a 
significant manner,
reduce the applicability of the strictly one-dimensional 
framework---at least for states 
in the low-energy regime characterized as gas-like three-atom states.

\section{Pancake-shaped trap}
\label{sec_pancake}
For pancake-shaped traps with $\eta < 1$,
we use the following form of the Green's function
$\mathcal{G}^{\rm{3D}}$~\cite{calarco-r,calarco},
\begin{align}
\label{eq_G_pancake}
\mathcal{G}^{\rm{3D}}(
[\epsilon_{\boldsymbol\lambda}+\eta+1/2]E_z
;{\bf{r}};{\bf{0}})=
\frac{1}{\sqrt{\pi} E_z a_z^3}\exp\left(-\frac{ z^2}{2 a_z^2}\right)
\sum_{j=0}^{\infty} 
\frac{(-1)^j}{2^{2j}j!}
H_{2j}(z/a_z) 
\mathcal{G}^{\rm{2D}}
\left( \left[
\frac{\epsilon_{\boldsymbol\lambda}-2 j}{\eta}+1 \right]E_{\rho};
\rho, 0\right).
\end{align}
This expression 
is equivalent to Eq.~(\ref{eq_G_cigar}) but converges faster for 
pancake-shaped traps than Eq.~(\ref{eq_G_cigar}).
Using Eq.~(\ref{eq_G_pancake}) in
Eq.~(\ref{eq_Ilambda}), we obtain
\begin{eqnarray}
\label{eq_Ilambda_pancake}
I_{{\boldsymbol{\lambda}},
\boldsymbol{\lambda'}}^{\rm{3D}}(\epsilon_{\boldsymbol{\lambda'}}) =
2 \sqrt{\pi}(-1)^m \delta_{m,m'}\times \nonumber \\
\lim_{j_{\rm{max}} \rightarrow \infty} \sum_{j=0}^{j_{\rm{max}}}
\frac{(-1)^j\sqrt{\pi^{1/2}(2j)!}}{2^{j} j!}
I_{n_z,n_z'}^{\rm{p}}(j)
I_{n_{\rho},n_{\rho}',m}^{\rm{p}}(\epsilon_{\boldsymbol{\lambda'}},j),
\end{eqnarray}
where
\begin{align}
\label{eq_Iz_pancake}
I_{n_z,n_z'}^{\rm{p}}(j)=
a_z^{1/2}
\int_{-\infty}^{\infty} \varphi_{2j}\left(\frac{\sqrt{3}}{2}z\right)
 \varphi_{n_z'}\left(-\frac{z}{2}\right) \varphi_{n_z}(z)~dz
\end{align}
and
\begin{align}
\label{eq_Irho_pancake}
&I_{n_{\rho},n_{\rho}',m}^{\rm{p}}(\epsilon_{\boldsymbol{\lambda'}},j)=
\nonumber \\
&E_z a_z^2 \int_0^{\infty} 
\mathcal{G}^{\rm{2D}}\left(
\left[\frac{\epsilon_{\boldsymbol\lambda'}-2 j}{\eta}+1
\right]E_{\rho};
\frac{\sqrt{3}}{2}\rho; 0\right)
R_{n_{\rho}',m}\left(\frac{\rho}{2}\right)
R_{n_{\rho},m}(\rho) 
~\rho d\rho.
\end{align}
Details regarding the evaluation of the integrals are 
explained in Appendix~\ref{appendix}.
In the following, we limit ourselves to cases where the reciprocal of the
aspect ratio is an integer.
In this case, we have~\cite{calarco-r,calarco}
\begin{align}
\label{eq_f41_pancake}
\mathcal{F}^{\rm{3D}}(\epsilon_{\boldsymbol{\lambda}},\eta)=
-2 \eta \sum_{k=0}^{1/\eta-1}
\frac{\Gamma(-\frac{\epsilon_{\boldsymbol{\lambda}}}{2} + k \eta)}{\Gamma(-\frac{\epsilon_{\boldsymbol{\lambda}} + 1}{2} + k \eta)}.
\end{align}

\begin{figure}
\vspace*{+.9cm}
\includegraphics[angle=0,width=70mm]{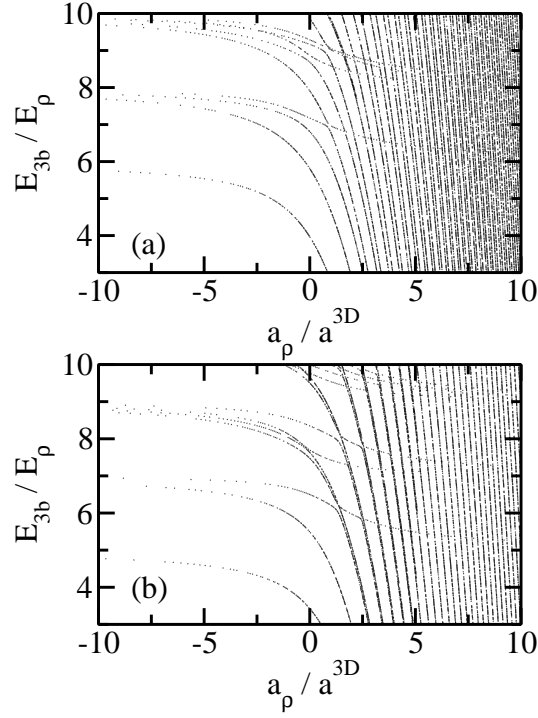}
\vspace*{0.2cm}
\caption{Relative three-body energies $E_{\rm{3b}}/E_{\rho}$ 
as a function of the 
inverse scattering length $a_{\rho}/a^{\rm{3D}}$ for a pancake-shaped
trap with aspect ratio $\eta = 1/2$ and 
(a) $M=0$ and $\Pi_z=+1$, and (b) $M=\pm 1$ and $\Pi_z=+1$.}
\label{fig_pancake}
\end{figure}

Figure~\ref{fig_pancake}
shows the relative three-body energies $E_{\rm{3b}}/E_{\rho}$ as a function of 
the inverse scattering length for $\eta = 1/2$, $\Pi_z=+1$,
and (a) $M=0$ and (b) $M=\pm1$.
In the $(a^{\rm{3D}})^{-1} \to -\infty$ limit, the ground state has 
$M=\pm 1$ and $\Pi_z=+1$ symmetry.
In the $(a^{\rm{3D}})^{-1} \to +\infty$ limit, in contrast,
the ground state has $M=0$ and $\Pi_z=+1$ symmetry.

\begin{figure}
\vspace*{+.9cm}
\includegraphics[angle=0,width=70mm]{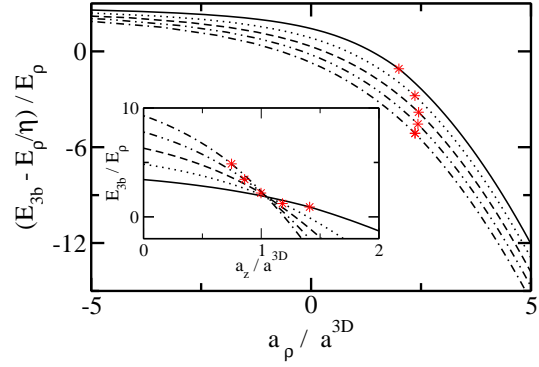}
\vspace*{0.5cm}
\caption{(Color online)~``Crossover curve'' of the 
three-body system with $\Pi_z=+1$, shifted by 
$E_z = E_{\rho} / \eta$, for various aspect ratios
of the trap, $\eta = 1/2$ (solid line), $1/4$ (dotted line), 
$1/6$ (dashed line), $1/8$ (dash-dot-dotted line), 
and $1/10$ (dash-dotted line) as a function of $a_{\rho} / a^{\rm{3D}}$.
The scattering lengths at which the $M$ quantum number of the corresponding 
eigenstate changes from 
$M = \pm 1$ (``left side of the graph'') to 
$M = 0$ (``right side of the graph'') are marked by asterisks.
The inset shows the (unshifted)
crossover curve as a function of $a_z / a^{\rm{3D}}$.}
\label{fig_pancake_gr_st}
\end{figure}

Figure~\ref{fig_pancake_gr_st}
shows the relative energy of the energetically 
lowest-lying state, the so-called crossover curve, 
of the three-body system with $\Pi_z=+1$ for various aspect ratios
of the trap ($\eta = 1/2, \cdots, 1/10$)
as a function of the inverse scattering length $a_{\rho}/a^{\rm{3D}}$. 
For comparative purposes, we subtract the relative ground state
energy of $E_z$ of the non-interacting one-dimensional system,
that is, the energy that the system would have in the
$z$-direction if the dynamics in the tight confinement 
direction were frozen,
from the full three-dimensional energy.
The scattering lengths at which the symmetry of the corresponding 
eigenstate changes from 
$M = \pm 1$ to 
$M = 0$ are marked by asterisks.
The symmetry change occurs around $a_z/a^{\rm{3D}} \approx 1$
(see inset).

It is instructive to compare Fig.~\ref{fig_pancake_gr_st}
(pancake-shaped trap) and Fig.~\ref{fig_cigar_gr_st}
(cigar-shaped trap).
For both geometries, the crossover curve changes symmetry.
The change of the symmetry is associated with the
low-energy coordinate (the $\rho$-coordinate for pancake-shaped systems
and the $z$-coordinate for cigar-shaped systems).
For both geometries,
the symmetry change occurs,
for the aspect ratios considered,
when $a^{\rm{3D}}$ is of the order of the
oscillator length in the tight confinement direction.

\begin{figure}
\vspace*{+.9cm}
\includegraphics[angle=0,width=70mm]{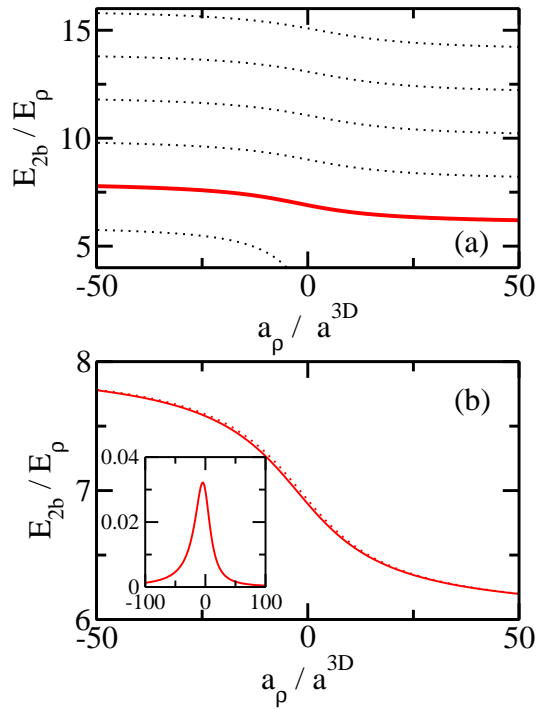}
\vspace*{0.2cm}
\caption{(Color online)~(a)~Relative two-body energies $E_{\rm{2b}}/E_{\rho}$ 
as a function of the 
inverse scattering length $a_{\rho}/a^{\rm{3D}}$ for a pancake-shaped
trap with aspect ratio $\eta = 1/10$, $M=0$ and $\Pi_z=+1$.
(b)~The solid curve from panel (a) is replotted and compared with the
energy obtained by solving the strictly two-dimensional 
eigenequation with renormalized 
two-dimensional scattering length $a_{\rm{ren}}^{\rm{2D}}$
(dotted line; 
the energy $E_z/2$ has been added to allow for a comparison 
with the full three-dimensional energy).
The inset shows the difference between the dotted and solid lines as 
a function of the inverse scattering length $a_{\rho}/a^{\rm{3D}}$.
The scale of the $y$-axis is identical to that of the inset 
of Fig.~\ref{fig_pancake_quasi_3-body}(b).}
\label{fig_pancake_quasi_2-body}
\end{figure}

Next, we consider the small $\eta$ limit in more detail.
For $\eta\ll 1$ and 
$|\epsilon|\ll 1$, we have~\cite{calarco-r,calarco}
\begin{align}
\label{eq_f_small_eta}
\left. \mathcal{F}^{\rm{3D}}(\epsilon, \eta)\right|_{\eta\ll 1}
\approx 
\frac{1}{\sqrt{\pi}}
\left[ 2 \mathcal{F}^{\rm{2D}}(\epsilon)
-2 \ln (\mathcal{C})
- \ln(\eta) 
\right],
\end{align}
and the two-body eigenequation for the relative energy becomes
\begin{eqnarray}
\label{eq_two-b_2d}
\mathcal{F}^{\rm{2D}}(\epsilon)= \ln(a^{\rm{2D}}_{\rm{ren}}/a_{\rho}),
\end{eqnarray}
 where the renormalized two-dimensional scattering length
$a_{\rm{ren}}^{\rm{2D}}$ is given by~\cite{petrov}
\begin{eqnarray}
\label{eq_ren_a_2d}
\frac{a_{\rm{ren}}^{\rm{2D}}}{a_{\rho}}= 
\sqrt{\eta}
{\mathcal{C}} \exp \left( -\frac{\sqrt{\pi}a_z}{2 a^{\rm{3D}}} \right)
\end{eqnarray}
with ${\mathcal{C}} \approx 1.479$.
Figure~\ref{fig_pancake_quasi_2-body}(a)
shows the relative two-body energies for a system with 
$\eta = 1/10$, $M=0$ and $\Pi_z=+1$ obtained by solving
the eigenequation
$\mathcal{F}^{\rm{3D}}(\epsilon, \eta=1/10)=-a_z/a^{\rm{3D}}$
[see Eq.~(\ref{eq_f41_pancake}) for $\mathcal{F}^{\rm{3D}}(\epsilon, \eta)$].
Figure~\ref{fig_pancake_quasi_2-body}(b)
compares the full three-dimensional energy (solid line)
with the energy obtained by solving the strictly 
two-dimensional eigenequation, Eq.~(\ref{eq_two-b_2d}),
with renormalized two-dimensional scattering length 
$a_{\rm{ren}}^{\rm{2D}}$ (dotted line).
For comparative purposes,
we add the energy of the tight 
confinement direction to the energy of the strictly two-dimensional system.
The inset of Fig.~\ref{fig_pancake_quasi_2-body}(b)
shows the difference between the strictly two-dimensional energy
and the full three-dimensional energy as a function of $a_{\rho}/a^{\rm{3D}}$.
The maximum deviation occurs around unitarity and is of the 
order of $0.4 \%$.

To treat the three-body system in the small $\eta$ limit,
we insert Eqs.~(\ref{eq_f_small_eta}) and~(\ref{eq_ren_a_2d})
into Eq.~(\ref{eq_eigen2}). This yields
\begin{align}
\label{eq_eigen_small_eta}
\sum_{\boldsymbol{\lambda'}}\left[
 \frac{\sqrt{\pi}}{2} I^{\rm{3D}}_{\boldsymbol{\lambda},\boldsymbol{\lambda'}}
(\epsilon_{\boldsymbol{\lambda}'})-
\mathcal{F}^{\rm{2D}}(\epsilon_{\boldsymbol{\lambda}'})
\delta_{\boldsymbol{\lambda},\boldsymbol{\lambda}'}
 \right]f_{\boldsymbol{\lambda}'}
= \ln \left(\frac{a_{\rho}}{ a_{\rm{ren}}^{\rm{2D}}} \right) f_{\boldsymbol{\lambda}}.
\end{align}
For fixed $M$, Eq.~(\ref{eq_eigen_small_eta})
reduces to the strictly two-dimensional eigenequation,
Eq.~(\ref{eq_eigen_2d}), if
{\em{(i)}} the sum over
$\boldsymbol{\lambda}'$ is restricted to a sum over $n_{\rho}'$ 
[i.e., if $\boldsymbol{\lambda}'=(0,n_{\rho}',m'=M)$];
{\em{(ii)}} the index $\boldsymbol{\lambda}$ is restricted to 
$\boldsymbol{\lambda}=(0,n_{\rho},m=M)$; 
{\em{(iii)}} the energy $E_{\rm{3b}}$ is replaced by $E_{\rm{3b}} -E_z$;
and
{\em{(iv)}} $j_{\rm{max}}$ in Eq.~(\ref{eq_Ilambda_pancake}) is set to zero.
Under these assumptions,  Eq.~(\ref{eq_eigen_small_eta})
reduces to Eq.~(\ref{eq_eigen_2d}) with $a^{\rm{2D}}$ 
replaced by $a_{\rm{ren}}^{\rm{2D}}$.

\begin{figure}
\vspace*{+.9cm}
\includegraphics[angle=0,width=70mm]{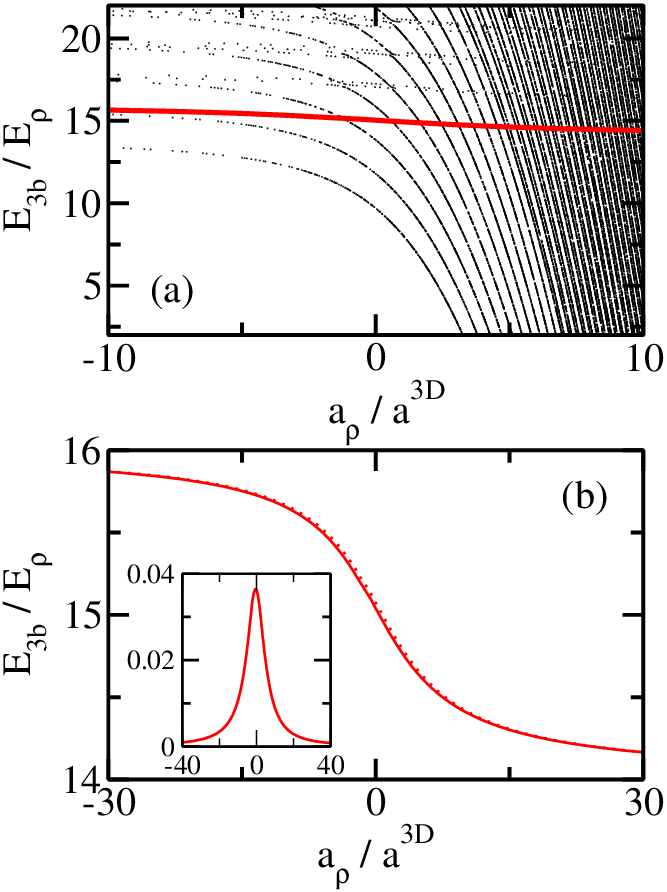}
\vspace*{0.2cm}
\caption{(Color online)~(a)~Relative three-body energies 
$E_{\rm{3b}}/E_{\rho}$ as a function of the 
inverse scattering length $a_{\rho}/a^{\rm{3D}}$ for a pancake-shaped
trap with aspect ratio $\eta = 1/10$, $M=0$ and $\Pi_z=+1$.
(b) The solid curve from panel (a) is replotted and compared with the
energy obtained by solving the strictly two-dimensional 
eigenequation with renormalized two-dimensional scattering
length $a_{\rm{ren}}^{\rm{2D}}$ (dotted line;
the energy of $E_{z}$ has been added 
to allow for a comparison with the full three-dimensional energy).
The inset shows the difference between the dotted and solid lines as 
a function of the inverse scattering length $a_{\rho}/a^{\rm{3D}}$.}
\label{fig_pancake_quasi_3-body}
\end{figure}

Figure~\ref{fig_pancake_quasi_3-body}(a)
shows the relative three-body energies for states with 
$M=0$ and $\Pi_z=+1$
as a function of the inverse scattering length $a_{\rho}/a^{\rm{3D}}$
for a pancake-shaped trap with $\eta=1/10$.
Figure~\ref{fig_pancake_quasi_3-body}(b)
compares the energy of the energetically lowest-lying
three-atom state with $M=0$ (solid line)
[see thick solid line in Fig.~\ref{fig_pancake_quasi_3-body}(a)]
with the corresponding state obtained by solving
the strictly two-dimensional equation with renormalized 
two-dimensional scattering length $a_{\rm{ren}}^{\rm{2D}}$
(dotted line).
The inset shows the difference between the energies obtained 
within the strictly two-dimensional and the full 
three-dimensional frameworks.
Similar to the one-dimensional case, the maximum of the deviation 
occurs near unitarity.
Comparison of the insets of Figs.~\ref{fig_pancake_quasi_2-body}(b)
and~\ref{fig_pancake_quasi_3-body}(b) suggests that,
at least in this low-energy example,
the presence of the third atom does not, in a 
significant manner,
reduce the applicability of the strictly two-dimensional 
framework.
For the same aspect ratio, 
the deviations are expected to increase with increasing
energy.

\section{Second- and third-order virial coefficients}
\label{sec_virial_coeff}
This section utilizes the two- and three-body energy spectra
to determine the second- and third-order virial coefficients 
as functions of the 
$s$-wave scattering length $a^{\rm{3D}}$, aspect ratio $\eta$ 
and temperature $T$.
The $n^{th}$-order virial coefficient $b_n$ enters into the
high-temperature expansion of 
the grand-canonical thermodynamic potential 
$\Omega$ of the equal-mass two-component
Fermi gas with interspecies $s$-wave 
interactions~\cite{ho-1,ho-2,rupak,drummond_prl,drummond_pra,drummond_2d,salomon-2010,zwierlein-2012,daily_2012},
$\Omega = \Omega^{(1)} + \Omega^{(2)} + \Omega^{(12)}$,
where $\Omega^{(1)}$ and $\Omega^{(2)}$ denote 
the grand-canonical thermodynamic potential of the spin-up component 
and the spin-down component, respectively, and $\Omega^{(12)}$ accounts
for the interspecies interactions,
\begin{align}
\label{eq_omega-12}
 \Omega^{(12)} = -k_{\rm{B}}T Q_{1,0} \sum_{n=2}^{\infty} b_n z^n.
\end{align}
Here, $z$ is the fugacity, $z=\exp[\mu/(k_{\rm{B}}T)]$.
In the high-temperature limit, $z$ is a small parameter and 
the expansion given in Eq.~(\ref{eq_omega-12}) is expected to provide
a good description if the sum is terminated at quadratic or cubic order.
The coefficient $b_2$ of the $z^2$ term is determined by one- 
and two-body physics and the coefficient $b_3$
of the $z^3$ term is determined by one-, two- and three-body 
physics. As the temperature approaches
the transition temperature from above, the de Broglie wave 
length increases and, correspondingly, the fugacity $z$
increases.
It follows that the number of $b_n$ coefficients needed to accurately
describe the thermodynamics increases
with decreasing temperature. Comparison with 
experimental data has shown
that the inclusion of $b_2$ and $b_3$ yields quite 
accurate descriptions of the high-temperature thermodynamics
of $s$-wave interacting two-component Fermi gases (see, e.g.,
Refs.~\cite{drummond_prl,salomon-2010,zwierlein-2012}).

In Eq.~(\ref{eq_omega-12}), $Q_{1,0}$
denotes the 
canonical partition function of a single spin-up particle.
We define the canonical partition function $Q_{n_1,n_2}$
of the system consisting of $n_1$ spin-up particles
and $n_2$ spin-down particles through
\begin{align}
\label{eq_can_parttion}
Q_{n_1,n_2} =  \sum_j \exp \left( -\frac{E_j^{\rm{tot}}(n_1,n_2)}{ k_{\rm{B}}T} \right),
\end{align}
where $E_j^{\rm{tot}}(n_1,n_2)$ denotes the total energy of the
system (including the center-of-mass energy)
and the summation over $j$ includes all quantum numbers 
allowed by the symmetry of the system.
For equal-mass fermions, as considered throughout this paper,
we have $Q_{1,0}=Q_{0,1}=Q_1$.
The virial coefficients $b_2$ and $b_3$ can be expressed 
as~\cite{drummond_prl,footnote1}
\begin{align}
\label{eq_b2_1}
b_2= \frac{Q_{1,1}-Q_1^2}{Q_1}
\end{align}
and
\begin{align}
\label{eq_b3_1}
b_3= 2~ \frac{Q_{2,1} - Q_{2,0} Q_1 - b_2 Q_1^2}{Q_1}.
\end{align}
The virial coefficients $b_2$ and $b_3$ depend on the interspecies 
scattering length $a^{\rm{3D}}$, aspect ratio $\eta$
and temperature $T$.
Once the thermodynamic potential is known, physical observables
such as the pressure and the entropy can be calculated.

We now discuss the determination of $b_2$ and $b_3$ for equal-mass
two-component Fermi gases under anisotropic harmonic confinement.
In this case, the single-particle canonical
partition function $Q_1$ can be determined analytically,
\begin{align}
\label{eq_1-body_can_partition}
Q_1 = \frac
{ \exp \left(\left[1/2+\eta \right]\tilde \omega_z \right)}
{\left[\exp \left(\tilde\omega_z \right) - 1\right] 
\left[\exp \left(\eta \tilde \omega_z \right)  - 1\right]^2},
\end{align}
where $\tilde \omega_z$ denotes the ``inverse temperature''
in units of $E_z$,
$\tilde \omega_z = E_z / (k_{\rm{B}}T)$.
Alternatively (see below), we
express the inverse temperature in units of $E_{\rho}$ or $E_{\rm{ave}}$,
$\tilde \omega_{\rho} = E_{\rho} / (k_{\rm{B}}T)$ and
$\tilde \omega_{\rm{ave}}(\eta) = E_{\rm{ave}} / (k_{\rm{B}}T)$.
The average energy $E_{\rm{ave}}$ is defined in terms of the 
root-mean-square or, in short, average
angular frequency 
$\omega_{\rm{ave}}(\eta)$, $E_{\rm{ave}}=\hbar \omega_{\rm{ave}}(\eta)$,
where
\begin{align}
\label{eq_omega-ave}
 \omega_{\rm{ave}}(\eta) = 
\sqrt{\frac {2 \omega_{\rho}^2 + \omega_z^2}{3}}.
\end{align}
We note that the average angular frequency coincides with the 
angular trapping frequency for $\eta=1$ but not for $\eta \ne 1$.
Below, we frequently suppress the explicit dependence 
of $\tilde{\omega}_{\rm{ave}}$ 
on $\eta$.
The partition functions $Q_{1,1}$ and $Q_{2,1}$ can be determined
from the two- and three-body energy spectra 
(see Secs.~\ref{sec_cigar} and~\ref{sec_pancake}) for 
each $s$-wave scattering length $a^{\rm{3D}}$, aspect ratio
$\eta$ and temperature $T$.

At unitarity, the high-temperature expansion of $b_n$ 
reads (for $\eta=1$, see Ref.~\cite{drummond_prl})
\begin{align}
\label{eq_bn_taylor}
b_n \approx b_n^{(0)} + b_n^{(2)}(\eta) \tilde \omega_{z}^2 
+ b_n^{(4)}(\eta) \tilde \omega_{z}^4 + \cdots.
\end{align}
The coefficients $b_n^{(0)}$, that is, the high-temperature limits 
of the trapped virial coefficients $b_n$, are independent of the 
aspect ratio $\eta$. 
This has previously been shown to be the case for $n=2$~\cite{peng_2011}.
Here, we extend the argument to all $n$.
Application of the local density 
approximation~\cite{menotti,drummond_prl,daily_2012} 
to axially symmetric confinement potentials shows that the virial
coefficients $b_n^{\rm{hom}}$ of the homogeneous system, 
which have been shown to be 
temperature-independent at 
unitarity~\cite{ho-1,ho-2,rupak}, 
 are related to $b_n^{(0)}$ through
\begin{align}
\label{eq_bn_hom}
b_n^{\rm{hom}} = n^{3/2} b_n^{(0)}.
\end{align}
Since Eq.~(\ref{eq_bn_hom}) holds for all $\eta$, $b_n^{(0)}$ 
has to be independent of $\eta$ for all $n$.

The expansion coefficients $b_n^{(k)}$, $k=2,4,\cdots$,
parametrize ``trap corrections'', that is, corrections 
that arise due to the fact that the harmonic confinement
defines a meaningful (finite) length scale.
In fact, for $\eta \neq 1$, the confinement defines two 
length scales, suggesting that the $b_n^{(k)}$, $k=2,4,\cdots$,
depend on $\eta$.
Equation~(\ref{eq_bn_taylor}) expresses the temperature
dependence of $b_n$ in terms of the 
inverse temperature associated with the $z$-direction,
regardless of whether $\eta$ is greater or smaller than 1.
Interestingly, it was shown in Ref.~\cite{peng_2011} that the 
dependence of $b_2^{(k)}$, $k=2,4,\cdots$, on the aspect ratio
can be parametrized, to a good approximation, in terms of the 
average inverse temperature $\tilde \omega_{\rm{ave}}(\eta)$,
\begin{align}
\label{eq_b2_approx}
b_2^{(k)}(\eta)\tilde \omega_z^{k} \approx 
b_2^{(k)}(1) [\tilde \omega_{\rm{ave}}(\eta)]^k.
\end{align}
Equation~(\ref{eq_b2_approx}) implies that the trap corrections 
for two-body systems with $\eta \ne 1$ can
be parametrized in terms of the trap corrections for the spherically
symmetric system if the inverse temperature is expressed in terms 
of the average trapping frequency that characterizes the 
anisotropic system.

\begin{figure}
\vspace*{+.9cm}
\includegraphics[angle=0,width=70mm]{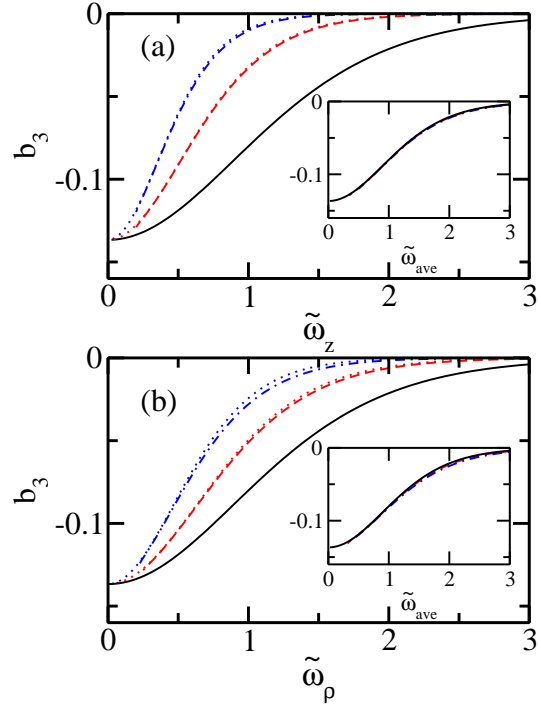}
\vspace*{0.2cm}
\caption{(Color online)~(a)~Third-order virial coefficient
$b_3$ at unitarity as a function of the 
inverse temperature $\tilde \omega_z$ for $\eta = 1$ (solid line), 
$2$ (dashed line), and $3$ (dash-dotted line). 
(b)~Third-order virial coefficient $b_3$  at unitarity as a 
function of the inverse temperature $\tilde \omega_{\rho}$ for 
$\eta = 1$ (solid line), $1/2$ (dashed line),
and $1/3$ (dash-dotted line). 
For $\eta \neq 1$, $b_3$ terminates at the inverse temperature
of about 0.25
since our calculations include a finite number of three-body energies;
obtaining the behavior of $b_3$ in the high-temperature limit
requires the inclusion of infinitely many three-body energies.
For 
$\eta \ne 1$,
dotted lines show $b_3$
for $\eta = 1$, calculated using the average
frequency of the respective anisotropic system.
This approximate description is quite good.
The insets of panels~(a) and (b)
show the same data as the main figure,
but now as a function of $\tilde{\omega}_{\rm{ave}}$
as opposed to $\tilde{\omega}_z$ and $\tilde{\omega}_{\rho}$.
The insets show that the third-order virial
coefficients for different $\eta$
collapse to a universal curve for all $\eta$
(deviations arise in the low-temperature regime,
i.e., for $\tilde{\omega}_{\rm{ave}} \gtrsim 1$).
}
\label{fig_virial-1}
\end{figure}

We now illustrate that Eq.~(\ref{eq_b2_approx}) applies 
not only to $b_2$ but also to $b_3$.
Figures~\ref{fig_virial-1}(a) and \ref{fig_virial-1}(b)
show the third-order virial coefficient
at unitarity for systems with $\eta \ge 1$ and 
$\eta \le 1$, respectively.
The virial coefficients are plotted as a function of the inverse
temperature expressed in units of the weak confinement direction,
i.e., in terms of $\omega_z$ for $\eta \ge 1$ and in terms of
$\omega_{\rho}$ for $\eta \le 1$.
In the high-temperature limit, $b_3$ approaches a constant, confirming
that $b_3^{(0)}$ is independent of $\eta$.
The dotted lines show the third-order virial coefficient
for $\eta=1$, calculated using the
average trapping frequency characteristic for the respective
anisotropic system.
Figure~\ref{fig_virial-1} illustrates that 
the third-order virial coefficient for anisotropic
traps is approximated well by that for $\eta=1$ with appropriately
scaled angular frequency.
The insets of Fig.~\ref{fig_virial-1} show that the third-order
virial coefficients of the anisotropic system collape, to a very
good approximation, to a universal curve over a surprisingly 
large temperature regime, i.e., down to temperatures
around $k_{\rm{B}} T \approx E_{\rm{ave}}/2$.
We conjecture that $b_n^{(k)}(\eta)\tilde \omega_z^{k}$ can be approximated
 quite well by $b_n^{(k)}(1) [\tilde \omega_{\rm{ave}}(\eta)]^k$ for 
$n=4,5,\cdots$ as well,
as long as $k$ is not 
too large, i.e, as long as the temperature is not too
low.

\begin{figure}
\vspace*{+.9cm}
\includegraphics[angle=0,width=70mm]{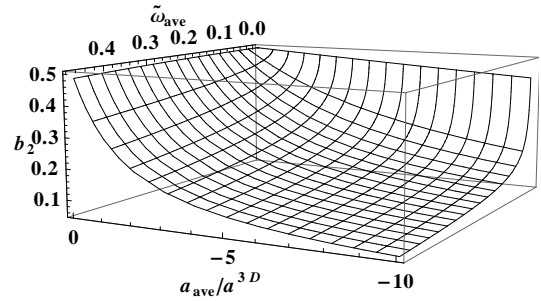}
\vspace*{0.2cm}
\caption{Second-order virial coefficient
$b_2$ for the two-body system
under isotropic confinement as a function of 
the inverse 
scattering length $a_{\rm{ave}}/a^{\rm{3D}}$ ($a^{\rm{3D}}$ negative)
and
the inverse 
temperature $\tilde{\omega}_{\rm{ave}}$.
The smallest $\tilde{\omega}_{\rm{ave}}$ considered
is $0.0003$. In the $\tilde{\omega}_{\rm{ave}} \rightarrow 0$
limit, $b_2$ approaches $1/2$ for all $a_{\rm{ave}}/a^{\rm{3D}}$
($a^{\rm{3D}}<0$; see text for further discussion).}
\label{fig_b2-diff-asc-3d}
\end{figure}

Next, we discuss the behavior of $b_2$ for finite $s$-wave scattering lengths.
Figure~\ref{fig_b2-diff-asc-3d}
shows a surface plot of $b_2$ for $\eta=1$ 
as a function of the inverse scattering length $a_{\rm{ave}}/a^{\rm{3D}}$
($a^{\rm{3D}} \leq 0$) and the inverse temperature $\tilde{\omega}_{\rm{ave}}$.
Here,
$a_{\rm{ave}}$ denotes the oscillator length associated with
the average trapping frequency,
$a_{\rm{ave}}=\sqrt{\hbar/(\mu \omega_{\rm{ave}})}$.
At unitarity, $b_2$ is only weakly-dependent on 
the temperature
and approximately equal to $1/2$ 
(see discussion above).
The smallest inverse temperature $\tilde{\omega}_{\rm{ave}}$
considered 
in Fig.~\ref{fig_b2-diff-asc-3d}
is $0.0003$. 
For this inverse temperature, $b_2$ is
fairly close to $1/2$ for all $a_{\rm{ave}}/a^{3D}$ shown.
Thus, Fig.~\ref{fig_b2-diff-asc-3d} shows that
the high-temperature limit of $b_2$ is 
nearly 
independent of 
the scattering length. 
This behavior can, as we now show, be understood from the two-body
energy spectrum.

\begin{figure}
\vspace*{+1.9cm}
\includegraphics[angle=0,width=70mm]{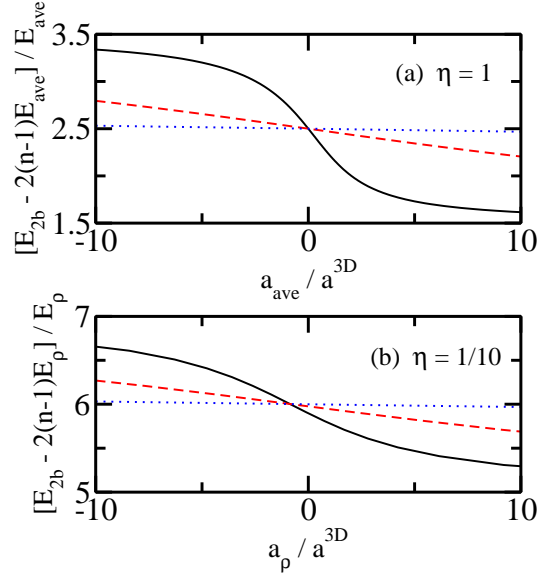}
\vspace*{2.3cm}
\caption{(Color online)~(a) 
Relative two-body energies $E_{\rm{2b}}/E_{\rm{ave}}$,
shifted by $2(n-1)$, 
for isotropic confinement
($\eta=1$ and $s$-wave channel)
as a function of 
the inverse scattering length $a_{\rm{ave}}/a^{\rm{3D}}$.
(b) Relative two-body energies $E_{\rm{2b}}/E_{\rho}$,
shifted by $2(n-1)$, 
for pancake-shaped confinement ($\eta=1/10, M=0$, and $\Pi_z=+1$)
as a function of 
the inverse scattering length $a_{\rho}/a^{\rm{3D}}$.
Solid, dotted and dashed lines show 
the energies
for $n=1, 100$ and $10000$, respectively.
}
\label{fig_2-body-E_1}
\end{figure}

Figure~\ref{fig_2-body-E_1}(a)
shows selected relative two-body energies as a function of 
$a_{\rm{ave}}/a^{\rm{3D}}$
for the trapped system with $\eta=1$.
In the low-energy regime (solid line), the two-body energy 
changes by nearly $2E_{\rm{ave}}$ for the scattering length
range shown.
As the energy increases (dashed and dotted lines show energies
around $200 E_{\rm{ave}}$ and 
$20000 E_{\rm{ave}}$, respectively), the two-body energy undergoes less of 
a change and eventually becomes nearly flat over the scattering length
region shown.
This implies that the high-energy portion of the two-body spectrum
looks like that of the unitary gas over an increasingly large region
around unitarity.
The behavior of the energy spectrum can be understood
by expanding the transcendental two-body eigenequation
$\mathcal{F}^{\rm{3D}}(\epsilon, 1)=-a_{\rm{ave}}/a^{\rm{3D}}$
around unitarity.
Assuming $|a_{\rm{ave}}/a^{\rm{3D}}| \ll 1$,
we find~\cite{peng_2011,footnotetypo}
\begin{align}
\label{eq_euni_taylor}
\frac{E_{\rm{2b}}}{E_{\rm{ave}}}  -
\left(2n+\frac{1}{2} \right) 
\approx  
-\frac{\Gamma(n+1/2)}
{\pi \Gamma(n+1)} \frac{a_{\rm{ave}}}{a^{\rm{3D}}}.
\end{align}
Equation~(\ref{eq_euni_taylor}) provides a good description
of the energies as long as the right hand side is small.
Since the right hand side of Eq.~(\ref{eq_euni_taylor}) scales
for large $n$
as $(a_{\rm{ave}}/a^{\rm{3D}}) / \sqrt{n}$, 
Eq.~(\ref{eq_euni_taylor})
provides, as $n$ increases, a good description for an increasingly
large region around unitarity.
That is, the energies vary approximately linearly with 
$a_{\rm{ave}}/a^{\rm{3D}}$, with a slope that approaches zero, as 
$n \to \infty$.
This analysis rationalizes why $b_2$
approaches $1/2$ as $T \rightarrow \infty$,
regardless of the value of the scattering length
($a^{\rm{3D}}<0$ and finite).

\begin{figure}
\vspace*{+.9cm}
\includegraphics[angle=0,width=70mm]{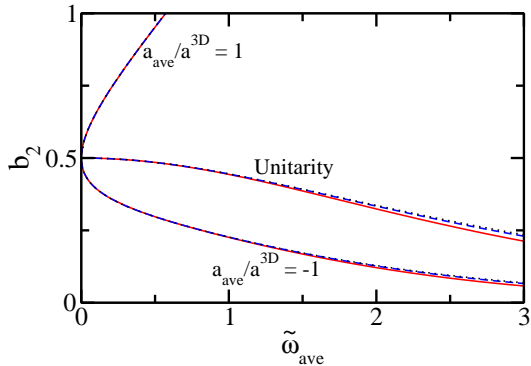}
\vspace*{0.4cm}
\caption{(Color online)~Second-order virial coefficient
$b_2$ of the trapped two-body system as a function of the inverse 
temperature $\tilde \omega_{\rm{ave}}$ for three different values of the inverse 
scattering length
($a_{\rm{ave}}/a^{\rm{3D}} = -1$, $0$ and $+1$; see labels)
for $\eta=1$ (solid lines), $\eta = 1/5$ (dashed lines) 
and $\eta = 1/100$ (dotted lines).}
\label{fig_b2-diff-asc}
\end{figure}

Figure~\ref{fig_b2-diff-asc-3d} has been obtained for a spherically
symmetric system, that is, for $\eta=1$.
We now demonstrate that Fig.~\ref{fig_b2-diff-asc-3d} applies,
for experimentally relevant temperatures,
to all aspect ratios and not just to $\eta=1$.
Figure~\ref{fig_b2-diff-asc}
shows $b_2$ as a function of $\tilde{\omega}_{\rm{ave}}$ 
for three different aspect ratios,
$\eta=1$ (solid lines), $\eta = 1/5$ (dashed lines) 
and $\eta = 1/100$ (dotted lines).
Three different scattering lengths are considered:
$a^{\rm{3D}} = -a_{\rm{ave}}$, $\infty$ and $a_{\rm{ave}}$.
It can be seen that $b_2$ is, to a very good approximation,
independent of $\eta$ in the high-temperature 
(small $\tilde{\omega}_{\rm{ave}}$) limit.
We stress that the independence of $b_2$ of $\eta$
requires that the inverse temperature and scattering length 
are expressed in terms of the 
average oscillator units $E_{\rm{ave}}$ and $a_{\rm{ave}}$, respectively.

To understand the universality implied by
Fig.~\ref{fig_b2-diff-asc}, that is, the fact that 
Fig.~\ref{fig_b2-diff-asc-3d} applies
to all aspect ratios and not just to $\eta=1$, we analyze
the behavior of the high-lying part of the relative 
two-body spectra for $\eta \ne 1$.
Figure~\ref{fig_2-body-E_1}(b) exemplarily shows the relative energies
for a pancake-shaped system with $\eta=1/10$.
Comparison with Fig.~\ref{fig_2-body-E_1}(a) shows that the 
qualitative behavior of the high-energy part of the spectrum
is independent of $\eta$.
This is confirmed by a more quantitative analysis that Taylor expands
the implicit eigenequation for the anisotropic two-body
system around unitarity.
We conclude that two two-body systems with different
aspect ratios but identical 
$a_{\rm{ave}}/a^{\rm{3D}}$ and $\tilde \omega_{\rm{ave}}$ are
characterized by approximately the same 
$b_2$ value. 
Our analysis of the three-body energies for
anisotropic confinement suggests that analogous
conclusions hold for $b_3$.
We speculate that the conclusions hold also for the virial coefficients
with $n>3$.

To estimate the extent of the universal behavior, it is instructive
to reexpress $\tilde{\omega}_{\rm{ave}}$ in terms of the Fermi temperature.
For a spin-balanced system of $N$ fermions under spherically 
symmetric confinement, we use the semi-classical expression 
$k_{\rm{B}} T_{\rm{F}}=(3 N)^{1/3} \hbar \omega_{\rm{ave}}$, yielding
$\tilde \omega_{\rm{ave}}=(3 N)^{-1/3}(T/T_{\rm{F}})^{-1}$.
For $N=10^2$, $10^4$ and $10^6$, $T/T_{\rm{F}}=1$ corresponds
to $\tilde \omega_{\rm{ave}} \approx 0.149$, 
$0.032$ and $0.007$, respectively.
For these temperatures, the thermodynamic behavior is,
according to our discussion above, expected to be essentially
universal over a fairly wide range of scattering lengths.
For $T/T_{\rm{F}}=1$,
we estimate that the deviation of $b_2$ from the value $1/2$
approaches
$5\%$ for 
$a_{\rm{ave}}/a^{\rm{3D}} \approx -0.16$, $-0.38$ and $-0.78$
for $N=10^2$, $10^4$ and $10^6$, respectively.
This implies that uncertainties 
of the scattering length
dependence on the magnetic field 
in recent experiments at
unitarity~\cite{salomon-2010,zwierlein-2012} should have a negligibly small
effect on the equation of state at unitarity.
For all three cases considered above,
the corresponding $(k_{\rm{F}} a^{\rm{3D}})^{-1}$ values 
is approximately $-0.03$.

The determination of the high-temperature behavior of the third-order
virial coefficient
for different scattering lengths and aspect ratios 
is much more demanding than that of the second-order virial coefficient.
Although our analysis of $b_3$ is less exhaustive than that of $b_2$,
it suggests that the conclusions drawn above for the 
second-order virial coefficient
carry over to the third-order virial coefficient.


\section{Conclusion}
\label{sec_conclusion}
This paper developed a Lippmann-Schwinger equation based approach 
to determine the energy spectrum and corresponding eigenstates
of three-body systems with zero-range $s$-wave interactions under
harmonic confinement with different transverse and
longitudinal angular trapping frequencies.
The formalism was applied to the equal-mass
system consisting of two identical fermions
and a third distinguishable particle in a different spin-state.
The energy spectra were determined as a function of the interspecies 
$s$-wave scattering length for various aspect ratios $\eta$,
$\eta>1$ (cigar-shaped trap) and $\eta<1$ (pancake-shaped trap).
For $\eta \gg 1$, we showed that the low-energy portion of the
energy spectra are reproduced well
by an effective one-dimensional Hamiltonian with renormalized 
one-dimensional coupling constant.
Similarly,
for $\eta \ll 1$, we showed that the low-energy portion of the
energy spectra are reproduced well
by an effective two-dimensional Hamiltonian with renormalized 
two-dimensional coupling constant.
As the energy increases, the description based on 
these effective low-dimensional
Hamiltonian deteriorates.

The two- and three-body energy spectra were then used to determine the second- and third-order virial
coefficients that determine the virial equation of state,
applicable to two-component Fermi gases at temperatures above the
Fermi temperature.
Our key findings are:

{\em{(i)}}
At unitarity, the second- and third-order virial coefficients
$b_2$ and $b_3$ approach constants in the high-temperature regime.
The constants (referred to as $b_2^{(0)}$ and $b_3^{(0)}$) are independent
of $\eta$.

{\em{(ii)}}
For finite scattering length $a^{\rm{3D}}$, we find that $b_2$ and $b_3$ 
collapse, to a very good approximation, to a single curve for all $\eta$
if the temperature and scattering length are expressed in terms of the 
average energy $E_{\rm{ave}}$ and 
the average oscillator length $a_{\rm{ave}}$,
respectively.
Deviations from the universal curve arise 
in the low-temperature regime where the virial
equation of state is not applicable.

{\em{(iii)}}
The virial coefficient $b_2$ is approximately equal to $1/2$ over a fairly 
large temperature and scattering length regime around unitarity.

The work presented in this paper is directly relevant to on-going 
cold atom experiments. The 
three-body spectra, e.g., can be 
measured experimentally through rf 
spectroscopy~\cite{esslinger,selim1,selim2}. 
Moreover, the
formalism can be employed to characterize the molecular states
in more detail, quantifying the ``perturbation'' of the
dimer due to the third particle throughout the dimensional crossover.
The determination of the virial coefficients is of 
immediate relevance to cold atom experiments that study the dynamics
of large fermionic clouds under  low-dimensional 
confinement.
The formalism developed in 
Secs.~\ref{sec_formalsolution}-\ref{sec_pancake}
of this paper can be extended fairly
straightforwardly to three-boson and unequal-mass systems.


\section{Acknowledgement}
We gratefully acknowledge support by the ARO and thank 
Krittika Goyal for contributions at the initial stage of this work.
This work was additionally supported by the National Science Foundation
through a grant for the Institute for Theoretical
Atomic, Molecular and Optical Physics at Harvard University
and Smithsonian Astrophysical Observatory.

\appendix
\section{Calculation of integrals involved}
\label{appendix}
This appendix presents the evaluation of the integrals defined in 
Secs.~\ref{sec_cigar} and~\ref{sec_pancake}.
The integrals 
$I_{n_{\rho},n_{\rho}',m}^{\rm{c}}(j)$ and $ I_{n_z,n_z'}^{\rm{p}}(j)$
are energy-independent. They are tabulated once and then used for each 
three-body energy considered.
To evaluate the integral 
$I_{n_{\rho},n_{\rho}',m}^{\rm{c}}(j)$, Eq.~(\ref{eq_Irho_cigar}),
we expand each of the three radial two-dimensional 
harmonic oscillator functions $R$ 
[see Eq.~(\ref{eq_two-d_wavefunction})]
in terms of a finite sum, i.e., we use the series expansion of the 
associated Laguerre polynomials~\cite{abramowitz},
\begin{align}
\label{eq_laguerre_series}
L_n^{(k)}(x)=
\sum_{j=0}^n(-1)^j \frac{(n+k)!}{(n-j)!(k+j)!j!} x^j.
\end{align}
$I_{n_{\rho},n_{\rho}',m}^{\rm{c}}(j)$ then becomes a finite triple sum 
that contains integrals of the form
\begin{align}
\label{eq_gaussian_integral}
\int_0^{\infty}\exp \left(-\frac{\eta \rho^2}{a_z^2} \right)
\left(\frac{\rho}{a_z} \right)^k ~d\rho =
\frac{1}{2} 
\left( \frac{1}{a_z} \right) ^k
\left( \frac{\eta}{a_z^2} \right)^{-\frac{1+k}{2}} 
\Gamma\left(\frac{1+k}{2}\right).
\end{align}
The finite sum is readily evaluated and 
$I_{n_{\rho},n_{\rho}',m}^{\rm{c}}(j)$ is stored for each 
$n_{\rho}$, $n_{\rho}'$, $m$ and $j$ combination.
To evaluate $ I_{n_z,n_z'}^{\rm{p}}(j)$, Eq.~(\ref{eq_Iz_pancake}),
we rewrite each of the three one-dimensional harmonic oscillator 
functions $\varphi$
[see Eq.~(\ref{eq_one-d_wavefunction})]
in terms of associated Laguerre polynomials instead of Hermite
polynomials~\cite{abramowitz_hermite},
\begin{align}
\label{eq_laguerre_hermit_1}
H_{2n}(z/a_z)=(-1)^n2^{2n}n! L_n^{(-1/2)}(z^2/a_z^2)
\end{align}
and
\begin{align}
\label{eq_laguerre_hermit_2}
H_{2n+1}(z/a_z)=(-1)^n 2^{2n+1} n! (z/a_z) L_n^{(1/2)}(z^2/a_z^2).
\end{align}
The evaluation of $ I_{n_z,n_z'}^{\rm{p}}(j)$ then proceeds
analogously to that of $I_{n_{\rho},n_{\rho}',m}^{\rm{c}}(j)$.

The integrals $I_{n_z,n_z'}^{\rm{c}}(\epsilon_{\boldsymbol{\lambda'}},j)$ and
$I_{n_{\rho},n_{\rho}',m}^{\rm{p}}(\epsilon_{\boldsymbol{\lambda'}},j)$,
Eqs.~(\ref{eq_Iz_cigar}) and~(\ref{eq_Irho_pancake}),
are energy-dependent.
To evaluate these integrals, we first ``separate out'' 
the energy dependence and then proceed along the lines discussed above.
The energy dependence of $I_{n_z,n_z'}^{\rm{c}}(\epsilon_{\boldsymbol{\lambda'}},j)$
enters through 
$\mathcal{G}^{\rm{1D}}([\epsilon_{\boldsymbol\lambda'}-2 \eta j+1/2]E_z;
\sqrt{3}/2z; 0)$.
Using the identity~\cite{krittika_2007}
\begin{align}
\label{eq_hyper_series}
\Gamma(a)U\left(a,1/2,x\right)=\sum_{k=0}^{\infty}\frac{L_{k}^{(-1/2)}(x)}{k+a},
\end{align}
$I_{n_z,n_z'}^{\rm{c}}(\epsilon_{\boldsymbol{\lambda'}},j)$
can be written as an infinite series,
\begin{align}
\label{eq_Iz_series_cigar}
I_{n_z,n_z'}^{\rm{c}}(\epsilon_{\boldsymbol{\lambda'}},j)
= \lim_{k_{\rm{max}} \to \infty}
\frac{1}{2 \sqrt{2}}
\sum_{k=0}^{k_{\rm{max}}}\frac{1}{k-\frac{\epsilon_{\boldsymbol{\lambda'}} - 2 \eta j}{2}}
C_{n_z,n_z'}^k,
\end{align}
where the coefficients $C_{n_z,n_z'}^k$ are energy-independent,
\begin{align}
\label{eq_Iz_cnnpk}
C_{n_z,n_z'}^k = \int_{-\infty}^{\infty} 
\varphi_{n_z}(z) \varphi_{n_z'}(-z/2) 
\exp{\left(-\frac{3z^2}{8a_z^2} \right) }
L_{k}^{(-1/2)}\left(\frac{3z^2}{4 a_z^2}\right)dz.
\end{align}
The evaluation of the $C_{n_z,n_z'}^k$ now proceeds as above.

\begin{figure}
\vspace*{+.9cm}
\includegraphics[angle=0,width=70mm]{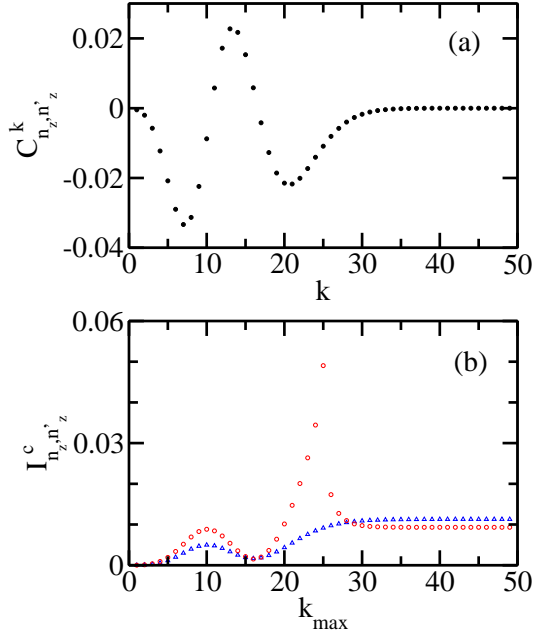}
\vspace*{0.2cm}
\caption{Convergence study.
(a) The dots show the coefficient $C_{n_z,n_z'}^k $, 
Eq.~(\ref{eq_Iz_cnnpk}), for 
$n_z=2$ and $n_z'=80$ as a function of $k$.
(b) The circles and triangles show the sum
$I_{n_z,n_z'}^{\rm{c}}(\epsilon_{\boldsymbol{\lambda'}},j)$,
Eq.~(\ref{eq_Iz_series_cigar}), for 
$(\epsilon_{\boldsymbol{\lambda'}} - 2 \eta j)/2= 25.745$
and $40.234$, 
respectively, as a function of the cutoff $k_{\rm{max}}$. 
As in panel~(a), we used $n_z=2$ and $n_z'=80$.}
\label{fig_conv}
\end{figure}

The series introduced in Eq.~(\ref{eq_Iz_series_cigar})
diverges if $(\epsilon_{\boldsymbol{\lambda'}} - 2 \eta j)/2$ 
is a positive integer or zero, that is, in the non-interacting limit.
In practice, this does not pose a constraint 
since the energy grid can be chosen such that 
it does not contain the non-interacting energies.
To analyze the dependence of 
$I_{n_z,n_z'}^{\rm{c}}(\epsilon_{\boldsymbol{\lambda'}},j)$
on the upper summation index $k_{\rm{max}}$, Fig.~\ref{fig_conv}(a)
shows the behavior of $C_{n_z,n_z'}^k$ for fixed $n_z$ and $n_z'$
as a function of $k$. In this example, $C_{n_z,n_z'}^k$  
vanishes nearly identically for $k \gtrsim 30$.
The coefficients $C_{n_z,n_z'}^k$ are multiplied by the $k$-dependent 
prefactor $[k-(\epsilon_{\boldsymbol{\lambda'}} - 2 \eta j)/2]^{-1}$.
This prefactor is maximal for
$k \approx (\epsilon_{\boldsymbol{\lambda'}} - 2 \eta j)/2$.
Figure~\ref{fig_conv}(b) shows the quantity 
$I_{n_z,n_z'}^{\rm{c}}(\epsilon_{\boldsymbol{\lambda'}},j)$
as a function of $k_{\rm{max}}$ for the same $n_z$ and $n_z'$ as in 
Fig.~\ref{fig_conv}(a) but two different values of 
$(\epsilon_{\boldsymbol{\lambda'}} - 2 \eta j)/2$,
that is, for
$(\epsilon_{\boldsymbol{\lambda'}} - 2 \eta j)/2=25.745$ and $40.234$.
In the first case, the 
absolute value of the prefactor takes its maximum at $k = 26$, 
where the coefficients $C_{n_z,n_z'}^k$ are non-zero.
Correspondingly, $I_{n_z,n_z'}^{\rm{c}}$ shows a sharp peak near 
$k_{\rm{max}} = 26$ and then smoothly approaches its asymptotic value
[see circles in Fig.~\ref{fig_conv}(b)]. 
In the second case, the absolute value
of the prefactor takes its maximum at 
$k = 40$,
where the $C_{n_z,n_z'}^k$ coefficients are very small. 
Correspondingly, the $C_{n_z,n_z'}^k$ coefficients suppress
the maximum of the prefactor and the quantity $I_{n_z,n_z'}^{\rm{c}}$
approaches its asymptotic value for $k_{\rm{max}} \approx 30$
[see triangles in Fig.~\ref{fig_conv}(b)]. 
We find that the choice of $k_{\rm{max}} \gtrsim 2 \max(n_z, n_z')$
yields well converged values for  
$I_{n_z,n_z'}^{\rm{c}}(\epsilon_{\boldsymbol{\lambda'}},j)$
for all energies considered.

In an alternative approach, we determined 
$I_{n_z,n_z'}^{\rm{c}}(\epsilon_{\boldsymbol{\lambda'}},j)$
through numerical integration
for each
$n_z$, $n_z'$, $n_{\rho}'$, $m$, $j$ and energy.
We have found this approach to be computationally more time-consuming than the 
tabulation approach discussed above.

The evaluation of the integral 
$I_{n_{\rho},n_{\rho}',m}^{\rm{p}}(\epsilon_{\boldsymbol{\lambda'}},j)$,
Eq.~(\ref{eq_Irho_pancake}), proceeds analogously to that of
$I_{n_z,n_z'}^{\rm{c}}(\epsilon_{\boldsymbol{\lambda'}},j)$. 
We use~\cite{krittika_2007}
\begin{align}
\label{eq_hyper_series_2}
\Gamma(a)U(a,1,x)=\sum_{k=0}^{\infty}\frac{L_{k}(x)}{k+a}
\end{align}
to separate out the energy-dependence that enters through 
$\mathcal{G}^{\rm{2D}}$.
The integral is then written as
\begin{align}
\label{eq_Irho_series_pancake}
I_{n_{\rho},n_{\rho}',m}^{\rm{p}}(\epsilon_{\boldsymbol{\lambda'}},j)=
\lim_{k_{\rm{max}} \rightarrow \infty}
\frac{1}{2 \pi}\sum_{k=0}^{k_{\rm{max}}}
\frac{1}{k-\frac{\epsilon_{\boldsymbol{\lambda'}} - 2 j}{2\eta}}
C_{n_{\rho},n_{\rho}',m}^k,
\end{align}
where
\begin{align}
\label{eq_Irho_cnnpk}
 C_{n_{\rho},n_{\rho}',m}^k =
&\int_0^{\infty} 
R_{n_{\rho},m}(\rho) 
R_{n_{\rho}',m}\left(\frac{\rho}{2}\right) 
\exp \left( -\frac{3 \eta \rho^2}{8 a_z^2} \right)
L_{k}\left(\frac{3 \eta \rho^2}{4 a_z^2} \right)\rho d\rho.
\end{align}
We tabulate $C_{n_{\rho},n_{\rho}',m}^k$ and calculate
$I_{n_{\rho},n_{\rho}',m}^{\rm{p}}(\epsilon_{\boldsymbol{\lambda'}},j)$ 
for each $\epsilon_{\boldsymbol{\lambda'}}$ ``on the fly''.
The convergence behavior of 
$I_{n_{\rho},n_{\rho}',m}^{\rm{p}}(\epsilon_{\boldsymbol{\lambda'}},j)$
is similar to that of
$I_{n_z,n_z'}^{\rm{c}}(\epsilon_{\boldsymbol{\lambda'}},j)$.

\end{document}